\documentclass{aa}  

\usepackage{graphicx}
\usepackage{txfonts}
\usepackage{colortbl}
\usepackage{hyperref}
\hypersetup{colorlinks=true,linkcolor=[rgb]{1.,0.2,0.2},citecolor=[rgb]{0.1,0.1,1.},filecolor=[rgb]{0.7,0.2,0.2},urlcolor=[rgb]{0.7,0.2,0.2}}
\usepackage{amsmath}
\usepackage{amssymb}
\usepackage{natbib}
\usepackage{multirow}

\usepackage[table,xcdraw]{xcolor}
\usepackage{adjustbox}
\def \LGalaxies{\texttt{L-Galaxies}\,}
\def \SupEdd{\textit{Loud}}
\def \Boosted{\textit{Boosted}}
\def \Fiducial{\textit{Quiet}}

\usepackage{ulem}

\def \msun{\,\rm M_\odot}

\usepackage{color}
\definecolor{myorange}{rgb}{0.8, 0.3, 0.0}

\definecolor{mygreen}{rgb}{0.0, 0.398, 0.0}

\def\msun{\,\rm{M_\odot}}

\begin{document}

   \title{Low-redshift  LISA massive black hole binaries in light of the nHz gravitational wave background}
   \title{Connecting low-redshift  LISA massive black hole mergers to the nHz stochastic gravitational wave background}
   \author{David Izquierdo-Villalba$^{*1,2}$ \and Alberto Sesana$^{1,2}$    \and Monica Colpi$^{1,2}$ \and Daniele Spinoso$^3$ \and \\ Matteo Bonetti$^{1,2}$ \and Silvia Bonoli$^{4,5}$ \and Rosa Valiante$^{6,7}$ }
   \institute{$^{1}$ Dipartimento di Fisica ``G. Occhialini'', Universit\`{a} degli Studi di Milano-Bicocca, Piazza della Scienza 3, I-20126 Milano, Italy\\
    $^{2}$ INFN, Sezione di Milano-Bicocca, Piazza della Scienza 3, 20126 Milano, Italy\\
    $^{3}$ Department of Astronomy, MongManWai Building, Tsinghua University, Beijing 100084, China\\
    $^{4}$ Donostia International Physics Centre (DIPC), Paseo Manuel de Lardizabal 4, 20018 Donostia-San Sebastian, Spain\\
    $^{5}$ IKERBASQUE, Basque Foundation for Science, E-48013, Bilbao, Spain\\
    $^{6}$ INAF/Osservatorio Astronomico di Roma, Via Frascati 33, 00040 Monte Porzio Catone, Italy \\
    $^{7}$ INFN, Sezione Roma1, Dipartimento di Fisica, ``Sapienza'' Universit\`{a} di Roma, Piazzale Aldo Moro 2, 00185, Roma, Italy \\ \\
   \email{david.izquierdovillalba@unimib.it}}

   \date{Received; accepted}

 
  \abstract
   {
   Pulsar Timing Array (PTA) experiments worldwide recently reported evidence of a nHz stochastic gravitational wave background (sGWB) compatible with the existence of slowly inspiralling massive black hole (MBH) binaries (MBHBs). The shape of the signal contains valuable information about the evolution of $z<1$ MBHs above $\rm 10^8\, M_{\odot}$, suggesting a faster dynamical evolution of MBHBs towards the gravitational-wave-driven inspiral or a larger MBH growth than usually assumed. In this work, we investigate if the nHz sGWB could also provide constraints on the population of merging lower-mass MBHBs ($\rm {<}\,10^7 \, M_{\odot}$) detectable by LISA. To this end, we use the \texttt{L-Galaxies} semi-analytical model applied to the \texttt{Millennium} suite of simulations. We generate a population of MBHs compatible simultaneously with current electromagnetic and nHz sGWB constraints by including the possibility that, in favourable environments, MBHs can accrete gas beyond the Eddington limit. The predictions of the model show that the global (integrated up to high-$z$) LISA detection rate is {\it not} significantly affected when compared to a fiducial model whose nHz sGWB signal is ${\sim}\,2$ times smaller. In both cases, the global rate yields ${\sim}\,12 \, \rm yr^{-1}$ and is dominated by systems of $\rm 10^{5-6} \, M_{\odot}$. The main differences are limited to low-$z$ ($z<3$), high-mass ($\rm {>}\rm 10^6\, M_{\odot}$) LISA MBHBs. The model compatible with the latest PTA results predicts up to ${\sim}\,1.6$ times more detections, with a rate of ${\sim}1\rm yr^{-1}$. We find that these LISA MBHB systems have 50\% probability of shining with bolometric luminosities $>\,10^{43}\rm erg/s$. Hence, in case PTA results are confirmed and given the current MBH modeling, our findings suggest there will be higher chances to perform multimessenger studies with LISA MBHB than previously expected.

   }
   \keywords{Methods: numerical --- quasars: supermassive black holes -- Gravitational waves -- Galaxies: interactions }
   \titlerunning{Low-$z$ LISA systems in light of PTA sGWB}
    \authorrunning{Izquierdo-Villalba et al}
   \maketitle
%

\section{Introduction}

The large fraction of massive galaxies hosting massive black holes (MBHs) together with the hierarchical assembly of galaxies suggest that massive black hole binaries (MBHBs) are unavoidable products of galaxy evolution. The evolution of these objects is ruled by many different processes acting on different scales \citep{Begelman1980}. The formation of these systems and their coalescence within the Hubble time is ruled by several processes. Large-scale ($\rm {\sim}\,kpc$)  dynamical friction, acting after galaxy mergers, causes the sink of the MBHs towards the center of the newly formed galaxy \citep{Milosavljevic2001,Yu2002,Mayer2007,Callegari2009,Callegari2011b,Bortolas2022,Kunyang2022,2023LRR-ASTRO-LISA}. At smaller separations (${<}\, \rm pc$), the interaction with single stars and a putative circumbinary disk surrounding the two MBHs brings the two objects down to a distance where the emission of gravitational waves (GWs) lead to their final coalescence 
\citep{Quinlan1997,Sesana2006,Vasiliev2014,Sesana2015,Escala2004,Escala2005,Dotti2007,Cuadra2009,Biava2019,Bonetti2020,Franchini2021,Franchini2022}.\\

Gravitational waves (GWs) emitted by MBHBs during the late inspiral, merger and ringdown phase at the end of their lives are the main target of current and future experiments. On the one hand, the \textit{Laser Interferometer Space Antenna} \citep[LISA,][]{LISA2017}, planned to be launched in 2035, will detect GWs at $0.1\,{-}\,100$ mHz, probing MBHBs in the $10^4\,{-}\,10^7 \, \msun$ range. On the other hand, \textit{Pulsar Timing Array} (PTA) experiments seek the nHz GWs produced by cosmologically nearby ($z<1$) and slowly inspiraling more massive
(${>}\,10^8\, \msun$) MBHBs. Currently, five main PTA experiments are taking data: the \textit{European Pulsar Timing Array} \citep[EPTA,][]{Kramer2013,Desvignes2016}, the \textit{North American Nanohertz Observatory for Gravitational Waves} \citep[NANOGrav,][]{McLaughlin2013,Arzoumanian2015}, \textit{Parkes Pulsar Timing Array} \citep[PPTA,][]{Manchester2013,Reardon2016}, the \textit{Indian
Pulsar Timing Array} \citep[InPTA][]{Joshi2022} and \textit{Chinese Pulsar Timing Array} \citep[CPTA,][]{Lee2016} projects. The recent results reported by these collaborations provide evidence of a stochastic GW background (sGWB) with amplitude $A$ that ranges between $[1.7\,{-}\,3.2] \,{\times}\,10^{-15}$, at a reference frequency $f=$1 $\rm yr^{-1}$ \citep{Agazie2023,Antoniadis2023,Reardon2023,Xu2023}. Different theoretical studies have been carried out to interpret the nature of the sGWB. Despite the signal being compatible with a population of MBHBs \citep{InterpretationPaperEPTA2023,Agazie2023}, modern, sophisticated hydrodynamical simulations and semi-analytical models, aiming at reproducing a wide array of cosmological observations, tend to produce smaller sGWBs, with typical amplitudes of $A\,{\approx}\,1\,{\times}\,10^{-15}$
\citep[see e.g][]{Sesana2009,Kelley2016,Kelley2017,Bonetti2018,IzquierdoVillalba2021,Curylo2023,Li2024}. Specifically, \cite{InterpretationPaperEPTA2023} showed that state-of-the-art semi-analytical models require important changes to reach the current sGWB reported by PTA collaborations. For instance, a fast dynamical evolution for MBHBs, a rapid and larger mass growth of MBHs, and a bigger normalization in the scaling relations are fundamental requirements in semi-analytical models to reconcile theoretical predictions and current observational constraints. However, these requisites are not easy to reach unless breaking current observational constraints of the MBH population, such as the black hole mass function or quasar luminosity function \citep[see e.g][]{IzquierdoVillalba2021,SatoPolito2023}.\\


Due to its high amplitude, the signal reported by the PTA collaborations is already providing valuable information about the possible formation and evolution of ${>}\,10^8\,\msun$ MBHBs. In this paper, we investigate whether this signal has also implications for LISA. In particular, we study how the forecasts for LISA MBHBs with a potential observable electromagnetic counterpart are affected when a galaxy formation model is tuned to reproduce {\it both} the latest results on the nHz sGWB {\it and} the electromagnetic emission of MBHs. To this end, we make use of the \LGalaxies{} semi-analytical model which is a unique framework that includes at the same level of detail the physics involved in the assembly of galaxies, MBHs, and MBHBs \citep{Henriques2015,Henriques2020,Yates2021,IzquierdoVillalba2020,IzquierdoVillalba2021,Spinoso2022}. Under the assumption that the whole PTA signal is coming from a low-$z$ population of MBHBs, the merger trees of the large \texttt{Millennium} simulations \citep{Springel2005} are used to explore which conditions and model modifications are required in order to match the theoretical predictions about MBHs and MBHBs with current PTA and quasars/AGN electromagnetic constraints. The effect of the new modeling in the LISA MBHBs is explored by using the \texttt{Millennium-II} merger trees \citep{Boylan-Kolchin2006} whose resolution is adequate to model smaller galaxies and MBHs down to $\approx 10^{3-4}\,\msun$. \\


The paper is organized as follows: In Section~\ref{sec:SAM_DESCRIPTIONS} we summarize the main physics included in \texttt{L-Galaxies} to trace the formation and evolution of galaxies, MBHs and MBHBs. Furthermore, we show the difficulties of current galaxy formation models to reach the high level of nHz sGWB without mismatching the number density of active MBHs. In Section~\ref{sec:SuperEdd_Model} we introduce a model able to generate an sGWB compatible with the latest PTA results, and to reduce the tension seen in the quasar bolometric luminosity function. In Section~\ref{sec:LISA_effects} we explore the effect of the nHz signal on the population of LISA MBHBs. In Section~\ref{sec:Caveats} we underline several caveats of the model. Finally, in Section~\ref{sec:Conclusions} we summarize the main results of the paper. A Lambda Cold Dark Matter $(\Lambda$CDM) cosmology with parameters $\Omega_{\rm m} \,{=}\,0.315$, $\Omega_{\rm \Lambda}\,{=}\,0.685$, $\Omega_{\rm b}\,{=}\,0.045$, $\sigma_{8}\,{=}\,0.9$ and $h \, {=} \, \rm H_0/100\,{=}\,67.3/100\, \rm km\,s^{-1}\,Mpc^{-1}$ is adopted throughout the paper \citep{PlanckCollaboration2014}.

\section{\LGalaxies{} semi-analytical model} \label{sec:SAM_DESCRIPTIONS}
In this section, we present \LGalaxies{} semi-analytical model  \citep[SAM,][]{Guo2011,Henriques2015,Henriques2020,Yates2021}. In brief, \LGalaxies{} is a code that tracks the cosmological assembly of galaxies through a set of analytical equations solved along the assembly history of dark matter halos, as given by their respective merger tree. On top of this, the latest modifications presented in \cite{IzquierdoVillalba2020,IzquierdoVillalba2021} and \cite{Spinoso2022} enable \LGalaxies{} to trace the formation and evolution of single and binary MBHs. We stress that all the physics described here, together with the default values of free parameters constitutes the SAM \textit{fiducial} model tagged as \Fiducial{} model (see Table~\ref{table:Models_Parameters}). 

\subsection{Dark matter} \label{sec:NBodySimulations}
\LGalaxies{} is a flexible semi-analytical model run on top of different dark matter (DM) merger trees extracted from N-body DM-only simulations. In particular, its performance has been tested in the \texttt{Millennium} and \texttt{TNG-DARK} suit of simulations \citep[see e.g.][]{Henriques2015,Ayromlou2021}. In this work, we use the merger trees extracted from the \texttt{Millennium} (MS, \citealt{Springel2005}) and \texttt{Millennium-II} (MSII, \citealt{Boylan-Kolchin2009}) simulations whose minimum particle mass and large cosmological volumes allow to trace the assembly of halos in a broad mass range ($10^8\,{-}\,10^{14}\, \msun$). MS follows the cosmological evolution of $2160^3$ DM particles of mass $8.6\,{\times}\, 10^8\, \mathrm {M_{\odot}}/h$ inside a periodic box of 500 ${\rm Mpc}/h$ on a side, from $z\,{=}\,127$ to the present. MSII can be thought as a high-resolution version of the MS, as it follows the same number of particles with a  a mass resolution 125 times higher ($6.885\,{\times}\,10^6\,\mathrm{M_{\odot}}/h$) in a box 125 times smaller  ($\mathrm{100\,Mpc}/h$). MS and MSII were stored at 63 and 68 epochs or snapshots, respectively. All the structures formed in these simulations were found by applying friend-of-fiend and \texttt{SUBFIND} algorithms and arranged with the \texttt{L-HALOTREE} code in the so-called \textit{merger trees} \citep{Springel2001}. Given the coarse time resolution offered by the outputs of MS/MSII (snapshots are separated by $\approx$300 Myr), \LGalaxies{} performs an internal time interpolation of 5-50 Myr (depending on redshift) to improve the tracing of the baryonic physics involved in galaxy evolution. Finally, both simulations, originally run with the WMAP1 \& 2dFGRS concordance cosmology, were re-scaled with the procedure of \cite{AnguloandWhite2010} to match the cosmological parameters provided by Planck first-year data \citep{PlanckCollaboration2014}.


\subsection{Gas and stars} \label{sec:GasAndStars}

\LGalaxies{} includes a sophisticated galaxy formation model, which is able to track the evolution of the gas and stellar components of structures forming along the DM merger trees. As soon as a DM halo collapses, the model assigns to it an amount of baryons consistent with the cosmological baryonic fraction \citep{WhiteFrenk1991}. These baryons are initially distributed in a quasi-static hot gas atmosphere which is able to cool down at a rate that depends on the redshift and mass of the hosting DM halo \citep{Guo2011}. The gas that is cooled falls at the centre of the DM halo leading to the formation of a gas disk due to angular momentum conservation. Star formation events occur as soon as the gas disk exceeds a critical mass, giving rise to a stellar component distributed in a disk \citep{Croton2006a}. As a consequence of star formation, massive and short-lived stars explode as supernovae injecting energy and metals into the cold gas disk, reheating it, and eventually pushing it beyond the virial radius of the DM halo \citep{Guo2011}. The cold gas component of massive galaxies is also regulated by the radio-mode feedback of the central MBH, which efficiently reduces or even suppresses cooling flows \citep{Croton2006}. Galaxies can also trigger star formation events through interaction with companion satellites. These events are divided into major and minor mergers. The first ones take place between galaxies with baryonic masses differing by less than a factor of 2 and the final result is the transformation of the remnant galaxy into a pure bulge. On the other hand, minor mergers occur during more extreme mass ratios and they are able to trigger the formation of bulges without destroying the stellar disc of the most massive galaxy. 
Besides mergers, massive disk-dominated galaxies are allowed to develop galactic bulges through disk instability \citep{Efstathio1982}. Finally, \LGalaxies{} models large-scale effects such as ram pressure stripping or galaxy tidal disruption \citep{Henriques2010,Guo2011}.

\subsection{Massive black holes} \label{sec:MBHs}
In this section we briefly describe the main physics included in \LGalaxies{} to deal with MBHs.

\subsubsection{Formation of massive black holes}
\LGalaxies{} includes a refined physical model to follow the genesis of MBHs \citep{Spinoso2022}. Specifically, it tracks the spatial variations of metals and Lyman Werner radiation to account for the formation of massive seeds ($\rm 10^3\,{-}\,10^5\, M_{\odot}$) via the direct collapse of pristine massive gas clouds and the collapse of dense, nuclear stellar clusters originated by early star-formation episodes. On the other hand, the formation of light seeds ($\rm 10\,{-}\,100\, M_{\odot}$) after the explosion of the first generation of stars (also known as PopIII stars) is also accounted for by using a subgrid approach which takes as an input the results of \texttt{GQD} Press–Schechter based SAM \citep{Valiante2021}. Notice that this seeding model is only used in the MSII trees since their high halo resolution allows us to track self-consistently the genesis of the first MBHs. On the contrary, the mass resolution of MS merger trees hinders the possibility of using the seeding model presented in \cite{Spinoso2022}. Therefore, when employing the MS simulation, we assume that all the newly resolved halos (which have mass of $\rm {\sim}\,10^{10}\,M_{\odot}$), regardless of redshift, host central MBHs with a fixed mass of $10^4\,\msun$. We have checked that the specific value of the initial MBH mass in the MS merger trees has a marginal effect in the sGWB at nHz frequencies. The cause is the fact that PTA MBHBs are placed in massive galaxies that undergo an intense merger history, feeding MBHs with large amounts of gas that erase any memory of the initial seed mass in a few hundred Myrs.

\subsubsection{Growth of massive black holes in the \Fiducial{} model} \label{sec:MBHPhysics}
As soon as the MBH forms, different processes can trigger its growth. In particular, in \LGalaxies{} the MBH growth is divided into three different channels: \textit{cold gas accretion}, \textit{hot gas accretion}, and \textit{mergers} with other MBHs. Among these, the first one is the main driver of black hole growth at any redshift and is triggered by galaxy mergers and disk instability (DI) events. On the one hand, after a galaxy merger, a fraction of cold gas is accreted by the nuclear black hole:
\begin{equation}\label{eq:QuasarMode_Merger}
\rm   \Delta {M}_{BH}^{gas} \,{=}\,\mathit{f}_{BH}^{merger} (1+\mathit{z}_{merger})^{\alpha} \frac{m_{R}}{1 + (V_{BH}/V_{200})^2}\, M_{\rm gas},
\end{equation}
where $\rm m_{R}\,{\leq}\,1$ is the baryonic ratio of the two interacting galaxies, $\rm V_{200}$ the virial velocity of the host DM subhalo, $z_{\rm merger}$ the redshift of the galaxy merger, $\rm M_{\rm gas}$ the cold gas mass of the galaxy. $\rm \mathit{f}_{BH}^{\rm merger}$, $\rm V_{BH}$ and $\alpha$ are three adjustable parameters set to $\rm 280 \, km/s$, $0.02$ and $5/2$, respectively, in what we refer hereafter as \Fiducial{} model (see Table~\ref{table:Models_Parameters}). While $\rm \mathit{f}_{BH}^{\rm merger}$ and $\rm V_{BH}$ characterize the efficiency of mergers in making gas lose angular momentum and flow towards the galactic nucleus, $\alpha$ takes into account the fact that at high-$z$ galaxies are more compact \citep[][]{Mo1998,Shen2003,vanderWel2014,Lange2015} and thus, any high-$z$ merger event should be more efficient in bringing gas onto the MBHs compared to low-$z$ events \citep[see][for similar approaches]{Bonoli2009,IzquierdoVillalba2020,IzquierdoVillalba2021}.\\

On the other hand, during a disk instability, the black hole accretes an amount of cold gas proportional to the mass of stars that trigger the stellar disk instability \citep[see][]{IzquierdoVillalba2020}:
\begin{equation}\label{eq:QuasarMode_DI}
\rm    \Delta {M}_{BH}^{gas} \,{=}\, \mathit{f}_{BH}^{DI} (1+\mathit{z}_{DI})^{\alpha} \frac{\Delta M_{stars}^{DI}}{{1 + (V_{BH}/V_{200})^2}},
\end{equation}
where $\rm \mathit{z}_{DI}$ is the redshift at which the disk instability takes place and $\rm \mathit{f}_{BH}^{DI}$ is a free parameter that takes into account the gas accretion efficiency, set to $0.0015$. $\rm V_{BH}$ and $\alpha$ have the same value as in the case of mergers. As described in Eq.~\ref{eq:QuasarMode_Merger}, all the free parameters involved in Eq.~\ref{eq:QuasarMode_DI} try to capture the efficiency of disk instabilities in feeding with cold gas the nuclear parts of the galaxy.\\

After a galaxy merger or a disk instability has occurred at a time $t_0$, the cold gas available for accretion ($\rm \Delta M_{BH}^{gas}$) is assumed to settle in a reservoir around the black hole, $\rm M_{Res}$\footnote{Notice that $\rm M_{Res} \,{=}\Delta M_{BH}^{gas}$ is only satisfied if before the galaxy merger of disc instability the reservoir around the MBH was empty. On the contrary, $\rm M_{Res} \,{=}\, \Delta M_{BH}^{gas}\,{+} M_{gas}^{left-over}$ being $\rm M_{gas}^{left-over}$ the leftover gas inside the reservoir, accumulated trough prior mergers or disc instabilities and not consumed by the MBH by the time at which the new merger or disc instability takes place.}. Instead of instantaneous gas consumption, the model considers that the gas reservoir is progressively consumed through an Eddington-limited growth phase, followed by a second phase of low accretion rates \citep{Hopkins2005,Hopkins2006a,Marulli2006,Bonoli2009}. To characterize these two stages we introduce in \LGalaxies{} the parameter $f_{\rm Edd}$\footnote{Given the value of $f_{\rm Edd}$, the mass of the MBH ($\rm M_{BH}$) and the radiative and accretion efficiency ($\eta$ and $\epsilon$, respectively) at a time $t$, the subsequent growth of a MBH ($\delta t$) is expressed as $\rm M_{BH}(\mathit{t}\,{+}\,\delta \mathit{t})\,{=}\,M_{BH}(\mathit{t})\,e^{\, \mathit{f}_{Edd} \frac{1-\eta(\mathit{t})}{\epsilon(\mathit{t})}\frac{\delta t}{\mathit{t}_{Edd}}}$
where $\rm \mathit{t}_{Edd}\,{=}\,0.45\, \rm Gyr$. Notice that $\eta$ accounts for the fraction of rest mass energy released by accretion (which depends on the MBH spin ($a$), see e.g Figure 5 of \citealt{King2008}), and $\epsilon\,{\leq}\, \eta$ accounts for the fact that not all of the available energy is necessarily radiated (which depends on the accretion disc geometry). Following \cite{MerloniANDHeinz2008}, \LGalaxies{} assumes that at $f_{\rm Edd}\,{>}\,0.03$ (thin disc regimen) $\epsilon \,{=}\,\eta(a)$ whereas at $f_{\rm Edd}\,{\leq}\,0.03$ (advected dominated accretion regimen) $\epsilon \,{=}\,\eta(a) f_{\rm Edd}/0.03$ \citep[see][]{IzquierdoVillalba2020}.}, defined as the ratio between the bolometric ($\rm L_{bol}$) and the Eddington luminosity ($\rm L_{Edd}$):

\begin{equation} \label{eq:feed_Edd}
    f_{\rm Edd} (t) \rm \,{=}\, \left \{ \begin{matrix}
      1 & \mathrm{M_{BH}}(t) \, {\le}\, \mathrm{M_{E}}   \\ \\
    \frac{1}{\left[ 1 + ((\mathit{t}-t_0)/\mathit{t}_\mathit{Q})^{1/2}\right]^{2/\beta} } &  \mathrm{M_{BH}}(t) \, {>} \, \mathrm{M_{E}}   \\ \\
  \end{matrix}\right.
\end{equation}
The duration of the Eddington limited phase is determined by the value of $\mathrm{M_{E}}\,{=}\, \mathrm{M_{BH}}(t_0) + \mathcal{F}_{\rm Edd} \mathrm{M_{Res}}(t_0)$, which is the mass reached by the MBH after consuming a fraction $\mathcal{F}_{\rm Edd}$ of its gas reservoir. Following  \cite{IzquierdoVillalba2020} the value of $\mathcal{F}_{\rm Edd}$ is set to 0.7.  Note that if a galaxy undergoes a new merger or DI while the central MBH is still accreting mass from a previous event, the new cold gas driven around the MBH environment is added to the previous remnant $\rm M_{Res}$ and the growth re-starts under the new initial conditions. Finally, $t_Q$ gives the time-scale at which $f_{\rm Edd}$ decreases and is defined as $ t_Q \,{=}\, t_d\,\xi^{\beta}/(\beta \ln 10)$, with $t_d \,{=}\, 1.26{\times}10^8 \, \rm yr $, $\beta \,{=}\, 0.4$ and $\xi \,{=}\, 0.3$. The choice of these values is based on \cite{Hopkins2009} who showed that models of \textit{self-regulated} MBH growth require $0.3\,{<}\,\beta\,{<}\,0.8$ and $0.2\,{<}\,\xi\,{<}\,0.4$. We highlight that any change in the values of $\beta$ and $\xi$ in the interval suggested by \cite{Hopkins2009} has a small effect on our results since the bulk of the MBH growth happens during the Eddington-limited phase. Finally, during any of the events that make the MBH grow, \LGalaxies{} tracks the evolution of the black hole spin ($a$) in a self-consistent way. During gas accretion events, the model uses the approach presented in \cite{Dotti2013} and \cite{Sesana2014} which links the number of accretion events that spin-up or spin-down the MBH with the degree of coherent motion in the bulge. On the other hand, after an MBH coalescence the final spin is determined by the expression of \cite{BarausseANDRezzolla2009} where a distinction between wet and dry mergers is done to compute the alignment/anti-alignment between the two MBHs. For further details on the implementation of the spin model inside \LGalaxies{}, we refer the reader to \cite{IzquierdoVillalba2020}.\\

\begin{figure*}
\centering  
\includegraphics[width=2.0\columnwidth]{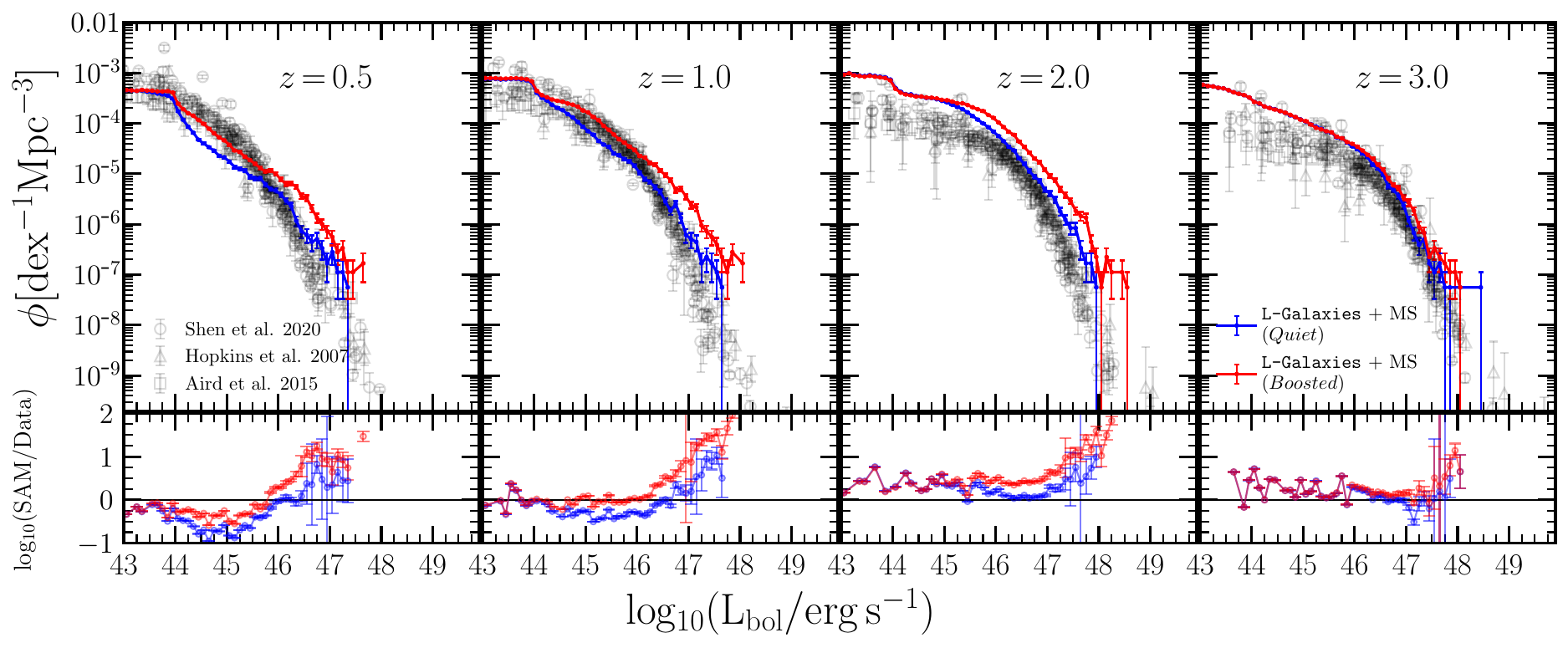}
\caption[]{Quasar luminosity functions at $z\,{=}\,0.5,1.0,2.0,3.0$ for the \Fiducial{} (blue) and \Boosted{} model (red) run in the \texttt{Millennium} merger trees (i.e \LGalaxies{} + MS). The error bars correspond to the Poissonian error. The results are compared with the observations of {\protect \cite{Hopkins2007} (triangles) \cite{Aird2015} (squares) and \cite{Shen2020} (circles)}. The lower panel represents the ratio between the model and the data points.}
\label{fig:LFs_Fiduvial_Fiducial_Boosted}
\end{figure*}

In Fig~\ref{fig:LFs_Fiduvial_Fiducial_Boosted} we compare the observational constraints of the quasar LFs \citep{Hopkins2007,Aird2015,Shen2020} with the predictions of the \Fiducial{} model of \LGalaxies{}. As we can see, the model is good agreement with the observations, being able to reproduce the redshift evolution of active MBHs from $z=0.5$ up to $z= 3$.

\renewcommand{\arraystretch}{1.1}
\begin{table}
\begin{adjustbox}{width=0.8\columnwidth,center} 
\begin{tabular}{lccc}
  Model &  $\mathit{f}_{\rm BH}^{\rm merger}$ & $\mathit{f}_{\rm BH}^{\rm DI}$ & $A(@1 \rm yr^{-1})$  \\ \hline 
{\Fiducial{} } &        0.020                    &   0.0015                       &            $1.2\,^{1.5}_{0.8}\,{\times}10^{-15}$      \\ \hline
{ \Boosted{} }           &    0.045                      &   0.0015                       &    $2.0\,^{2.8}_{1.2}\,{\times}10^{-15}$                 \\ \hline
{\SupEdd{} }    &     0.042                     &                     0.0015      &     $1.8\,^{2.3}_{1.2}\,{\times}10^{-15}$     \\ \hline
\end{tabular}
\end{adjustbox}
\caption{
Set of values assigned to the free parameters in every model. The errors on the sGWB amplitude have been computed by dividing the \texttt{Millennium} box into sub-boxes of $\rm 100 \, Mpc$ side-length. Then, it was computed with all of them the $\rm 16^{th}$ and $\rm 84^{th}$ percentile of $\rm A\,@1\,yr^{-1}$.}
\label{table:Models_Parameters}
\end{table}

\subsection{Massive black hole binaries} \label{sec:MBHBs}
The dynamical evolution of MBHB in \LGalaxies{} is divided into different stages \citep{Begelman1980}. The first one is called \textit{pairing phase} and it stars after the merger of the two galaxies (see Section~\ref{sec:GasAndStars}). During this process, the MBH hosted by the less massive galaxy undergoes dynamical friction which causes its sinking toward the galactic center of its new host. The process is modeled according to the standard \cite{BinneyTremine1987} equation which depends on the mass and orbital circularity of the MBH, the initial position at which the satellite galaxy deposited the MBH after the merger (${\sim}\,\rm kpc$ separation), and the velocity dispersion of the remnant galaxy. Notice that the model assumes that the nuclear MBH of the central galaxy is not displaced from the galactic center after the galactic merger. This is a fair assumption since the dynamical friction time scale is dominated by the lighter MBH making irrelevant the displacement of the central and most massive MBH. Finally, we stress that the secondary MBH might be embedded inside nuclear stellar clusters (NSCs, not accounted in the model yet) leading to larger effective masses during the dynamical fiction than the ones assumed in the MBHB model of \LGalaxies{}. This is a caveat that will be addressed in future works by using a phenomenological model (see e.g. \citealt{Polkas2023}) or by including a self-consisitent NSC modelling inside \LGalaxies{} (Hoyer et al. in prep).\\

Once the dynamical friction phase ends, the satellite MBH reaches the galactic nucleus of the new galaxy and it binds with the central MBH (${\sim}$ pc separation) starting the so-called hardening phase and giving rise to a massive black hole binary. While the most massive MBH of the binary is flagged as primary, the lighter one is tagged as secondary. The initial eccentricity of the binary orbit is randomly drawn in the range $[0,0.99]$, while the initial semi-major axis is set to the scale at which the stellar content of the galaxy (distributed according to a Sérsic model) equals the mass of the secondary MBH. The eccentricity and separation of the MHBH are then evolved self consistently according to the environment in which the system is embedded. In case the gas reservoir around the binary ($\rm M_{Res}$) is larger than its total mass ($\rm M_{Bin}$) the system evolves by interacting with a circumbinary gaseous disk following the prescription of \cite{Dotti2015}. Otherwise, the system is driven by the interaction with stars according to the theory developed by \cite{Quinlan1997,Sesana2015}. In both cases, GW emission takes over at smaller separations ($\rm {\lesssim}\,mpc$), leading the binary to final coalescence. We refer to \cite{IzquierdoVillalba2021} for a detailed description of the equations used to evolve the MBHB eccentricity and separation. In the case of repeated mergers, a third MBH can reach the nucleus of the remnant before the pre-existent binary completes its evolution. In this case, an MBH triplet forms and the outcome of the triple interaction is modeled according to the tabulated values of \cite{Bonetti2018ModelGrid}.\\



On top of the dynamical evolution of MBHBs, \LGalaxies{} allows MBHs in the pairing and hardening phase to increase their masses. Recent hydrodynamical simulations of merging galaxies with central MBHs have shown that the secondary galaxy suffers large perturbations during the pericenter passages around the central one \citep{Callegari2009,Callegari2011a,Callegari2011b,Capelo2015,Gabor2016}. Under these circumstances, the black hole of the secondary galaxy experiences accretion enhancements mainly correlated with the galaxy mass ratio. To include these findings, \LGalaxies{} assumes that right before the galaxy merger, the black hole of the secondary galaxy generates or increases its gas reservoir according to Eq.\ref{eq:QuasarMode_Merger}. Once the satellite MBH is deposited in the new galaxy and starts its pairing phase, the gas reservoir is consumed according to the two-phase model described in Section~\ref{sec:MBHPhysics}. The accretion process onto the pairing MBH lasts until it consumes the total gas reservoir stored before the merger. On the other hand, gas accretion onto MBHBs has been extensively explored by different theoretical studies \citep{DOrazio2013,Farris2014,Moody2019,Munoz2019,DOrazio2021}. The findings of these have shown that irrespective of the mass ratio of the binaries, the gas accretion onto the secondary MBH is sufficient to modify the final mass ratio of the binary, moving the initial values toward larger ones \citep[see e.g][]{Farris2014,Duffell2020}. Based on this picture, \LGalaxies{} assumes that an MBHB in the hardening phase featuring a gas reservoir progressively consumes it according to the results of \cite{Duffell2020}. In brief, the accretion rate of a primary black hole ($\dot{\rm M}_{\rm BH_1}$) is fully determined by the binary  mass ratio ($q$) and the accretion rate of the secondary black hole ($\dot{\rm M}_{\rm BH_2}$):
\begin{equation} \label{eq:Relation_accretion_hard_binary_blac_hole}
\dot{\rm M}_{\rm BH_1} =  \dot{\rm M}_{\rm BH_2} (0.1+0.9\mathit{q}),
\end{equation}
Except in the case of equal mass systems, secondary MBHs are farther from the binary centre of mass than primary ones. This causes them to be closer to the circumbinary disc edges and thus display high accretion rates. Based on this, \LGalaxies{} fix the accretion of the secondary black hole at the Eddington limit and determine the accretion onto the primary according to Eq.~\ref{eq:Relation_accretion_hard_binary_blac_hole}.

\subsection{Model constraints from Pulsar Timing Arrays} \label{sec:GalaxyFormationdiskrepancy}

Recent PTA results suggest the existence of an sGWB at nHz frequencies, compatible with a population of merging low-$z$ MBHBs \citep{Agazie2023,Antoniadis2023,Reardon2023,Xu2023}.\footnote{But see alternative origins of the sGWB from early cosmology (e.g \citealt{InterpretationPaperEPTA2023} and \citealt{Afzal2023} and references therein).} Despite the current significance is still below the canonical $5\sigma$ threshold, this signal provides galaxy formation models with a new condition to calibrate their underlying MBH growth physics, adding to the standard electromagnetic constraints coming from observations of galaxy properties and the quasar luminosity function.\\

The foundation to perform a comparison between recent PTA results and galaxy formation models resides in determining the comoving number density of MBHB mergers ($d^2n/dzd\mathcal{M}$) per unit redshift, $z$, and rest-frame chirp mass, $\mathcal{M}$\footnote{The chirp mass of an MBHB system is defined as ${\mathcal{M}}\,{=}\,\rm (M_{\rm BH,1}M_{\rm BH,2})^{3/5}(M_{\rm BH,1}\,{+}\,M_{BH,2})^{-1/5}$, being $\rm M_{\rm BH,1}$ and $\rm M_{\rm BH,2}$ the mass of the primary and secondary MBH, respectively}. Following \cite{Sesana2008} and making the specific assumption that inspiralling MBHBs in the PTA band are in circular orbits evolving purely due to GW emission, the characteristic sGWB can be written as:
\begin{equation} \label{eq:GWBG2}
    h^2_c(f) \, {=} \, \frac{4 G^{5/3} f^{-4/3}}{3c^2\pi^{1/3}} {\int}{\int} dz d\mathcal{M} \frac{d^2n}{dzd\mathcal{M}} \frac{\mathcal{M}^{5/3}}{(1+z)^{1/3}},
\end{equation}
where $f$ is the frequency of the GWs in the observer frame. This expression is often simplified as:
\begin{equation}
h_c(f) \,{=}\, A\left( \frac{f}{f_0} \right)^{-2/3} 
\end{equation}
where $A$ is the amplitude of the signal at the reference frequency $f_0$. From hereafter, we will set $f_0\,{=}\,1\,\rm yr^{-1}$ and refer $A$ to that frequency. Under these assumptions, the \Fiducial{} model of \LGalaxies{} predicts an $d^2n/dzd\mathcal{M}$ which generates a sGWB of amplitude $A \, {\sim}\,1.2\,{\times}\,10^{-15}$ (see Table~\ref{table:Models_Parameters}), fully ruled by $z\,{<}\,1$ MBHBs with $\mathcal{M}\,{>}\,10^8\, \msun$ \citep[see][]{IzquierdoVillalba2023}. Despite agreeing with past theoretical studies \citep{Wyithe2003,Sesana2008,Sesana2009,Sesana2013,2015MNRAS.447.2772R,Sesana2016,Bonetti2018ModelTriplets,Kelley2016,Siwek2020} its value is lower than the one reported by PTA collaborations which span over the range $[1.7\,{-}\,3.2] \,{\times}\,10^{-15}$ \citep{Antoniadis2023,Agazie2023,Reardon2023,Xu2023}. Interestingly, such tension is also present in other recent SAMs with MBHBs such as \cite{Li2024} or \cite{Curylo2023}.\\

To reconcile theoretical predictions with the recent PTA results, \cite{InterpretationPaperEPTA2023} explored changes in the dynamical models of MBHBs included in \LGalaxies{} (e.g. faster dynamical friction phases and only stellar hardening) showing that these modifications are not enough to reduce the discrepancy.  Despite this, in a recent paper, \cite{Barausse2023} showed that a Press–Schechter based SAM which includes a heavy MBH seeding scenario and assumes no delay between galaxy and MBH mergers would favor large sGWB, compatible with the latest PTA results.
Another approach was reported in \cite{IzquierdoVillalba2021} which proposed that MBHs should be more efficient in accreting cold gas after mergers and/or disk instabilities, thus becoming more massive when they enter the PTA frequency band. To increase the efficiency of MBH growth the authors explored in \LGalaxies{} a boosted model in which the two-phase growth model of Section~\ref{sec:MBHPhysics} was untouched but the gas accretion efficiency, $\rm \mathit{f}_{BH}^{merger}$ in Eq.~\eqref{eq:QuasarMode_Merger}, increases during mergers\footnote{The conclusions presented here apply also when increasing the gas accretion during disk instabilities.} (hereafter \Boosted{} model, see Table~\ref{table:Models_Parameters}). The results showed that the \Boosted{} model was able to generate a stochastic GW background of amplitude $[1.9\,{-}\,2.6]\,{\times}\,10^{-15}$, compatible with the latest PTA measurements. Despite this better agreement, the increase of the MBH growth was hindering the possibility of reproducing the number density of active MBHs. This is presented in  Fig.~\ref{fig:LFs_Fiduvial_Fiducial_Boosted} where we can see that the \Fiducial{} model is able to follow the observational trends presented in \cite{Hopkins2007}, \cite{Aird2015} and \cite{Shen2020}. However, the \Boosted{} one displays a systematic overprediction (up to ${\sim}\,2 \rm \, dex$) in the number density of bright quasars ($\rm L_{bol}\,{\gtrsim}\,10^{46} erg/s$) at $z\,{<}\,2$.\\

The results presented above suggest that increasing the mass of MBHs is a good avenue to reconcile theoretical models and PTA observations. However, the physical mechanism and the time scale by which the MBHs reach the necessary mass to generate a loud sGWB should be addressed carefully since the over-prediction of the quasar luminosity function seems to be a natural consequence. Taking this into account, in the next section we explore the effect of allowing super-Eddington accretion episodes in growing the population of single MBHs. Specifically, we investigate if these events enable the assembly of a population of MBHB compatible with the nHz sGWB and generate a population of active MBHs in agreement with the quasar luminosity functions. We stress that super-Eddington accretion is only enforced on single MBHs while the growth of MBHBs is modeled in the same way as described in Section~\ref{sec:MBHBs}.

\section{A faster assembly for the MBH population} \label{sec:SuperEdd_Model}

In this section, we explore the possibility of extending the \Fiducial{} model of \LGalaxies{} in such a way that galaxy mergers are more efficient in fuelling gas onto MBHs and \textit{some} MBHs, under certain conditions, can undergo super-critical accretion events. While the former requirement is done in the same way as in the \Boosted{} model (see Section~\ref{sec:GalaxyFormationdiskrepancy}), the latter is described in the next section. The interplay between these two processes is calibrated by running \LGalaxies{} on the \texttt{Millennium} merger trees. Finally, we stress again that super-Eddington accretion events will be allowed only to nuclear single MBHs, being the accretion onto MBHBs modelled in the same way as Section~\ref{sec:MBHBs}.

\subsection{A toy model for super-Eddington growth}

Super-Eddington accretion refers to growth episodes that proceed extremely rapidly, breaking the Eddington rate and increasing the MBH mass on time scales shorter than what is allowed by the standard Eddington-limited model \citep{Abramowicz1988}. Recent simulations have shown that not all the environments in which MBHs are embedded can trigger these extreme accretion events. Only dense and dusty gas environments around the MBHs replenished by large gas inflows after galactic mergers provide the ideal conditions to trigger super-Eddington growth \citep[see e.g][]{Inayoshi2016,Takeo2018,Regan2019,Toyouchi2021,Sassano2023,Massonneau2023}. To account for these requirements in the \LGalaxies{} SAM, we assume a super-Eddington phase is active only if: i) at the moment of the merger/disk instability ($t_0$) the gas reservoir around the single MBH exceeds by a factor of $\mathcal{R}^{\rm th}$ the MBH mass:
\begin{equation}
\rm \frac{M_{Res}(t_0)}{M_{BH}(t_0)}\,{>}\,\mathcal{R}^{th}\\     
\end{equation}
and ii) the rate at which the gas is infalling towards the galactic centre, $\rm M_{infolw}$, overcomes a certain threshold given by
\begin{equation}
\rm M_{inflow} = \frac{\Delta M_{BH}^{gas}}{\mathit{t}_{dyn}} \,{>}\, M_{inflow}^{th},     
\end{equation}
where $\rm \Delta M_{BH}^{gas}$ corresponds to the cold gas that fuels the new accretion event onto the nuclear single MBH and it is determined by Eq.~\eqref{eq:QuasarMode_Merger} and Eq.~\eqref{eq:QuasarMode_DI}. We set the dynamical time of the gas to inflow towards the galactic nucleus, $t_{\rm dyn},$, as $\rm V_{disk}/\mathit{R}_{gas}^{sl}$ being $\rm V_{disk}$ the maximum circular velocity of the cold gas (assuming an exponential disk profile) and $R_{\rm gas}^{\rm sl}$ the cold gas scale length radius \citep[see][for the detailed model of galactic sizes included in \LGalaxies{}]{Guo2011}.\\

In the case both conditions are satisfied, the MBH does not follow the Eddington-limit growth of Eq.~\eqref{eq:feed_Edd} but undergoes a super-critical accretion event whose lightcurve is characterized by the following $f_{\rm Edd}$:

\begin{equation} \label{eq:feed_SE}
    f_{\rm Edd} (t) \rm \,{=}\, \left \{ \begin{matrix}
    B(a) [ \frac{0.985}{\dot{\rm M}_{\rm Edd}/\dot{\rm M} \,{+}\, C(a)}  & \\
      \,\,\, \,\,\, \,\,\, \,\,\, \,\,\, \,\,\, +  \frac{0.015}{\dot{\rm M}_{\rm Edd}/\dot{\rm M} \,{+}\, D(a)} ] & \mathrm{M_{BH}}(t) {\le} \mathrm{M_{SE}} \\ \\
    \frac{1}{\left[ 1 + ((\mathit{t}-t_0)/\mathit{t}_\mathit{Q})^{1/2}\right]^{2/\beta} } &  \mathrm{M_{BH}}(t) {>} \mathrm{M_{SE}}   \\ \\
    \end{matrix}\right.
\end{equation}
%

\noindent This two-phase-lightcurve 
tries to mimic the fact that even in an environment favourable for super-Eddington accretion, the AGN feedback resulting from the gas accretion makes the MBH reach a self-regulated phase within a few Myr\footnote{We have checked that in the model the typical time spent by the MBH in the super-critical accretion is ${\sim}\,70\,{-}\,100\, \rm Myr$. Notice that the time resolution of the SAM is ${\sim}\,20\,\rm Myr$. These values align with the results of \cite{Pezzulli2016} and \cite{Lupi2023} which pointed out that super-Eddington accretion events can be sustained over time scales of a few tens of Myrs.} \citep[see e.g][]{Massonneau2023}. The parameter $\rm M_{SE}$ is the maximum mass reached by the MBH during the super-critical accretion and it is defined as $ \mathrm{M_{SE}} \,{=} \, \mathrm{M_{BH}} (t_0) + \mathcal{F}_{\rm SE} \mathrm{M_{Res}} (t_0)$, being $\mathrm{M_{BH}} (t_0)$ and $\mathrm{M_{Res}} (t_0)$ the mass of the MBH and the reservoir at the moment of the (major/minor) merger and/or disk instability (occurring at $t_0$). $\mathcal{F}_{\rm SE} $ determines the fraction of the gas reservoir consumed by the MBH throughout the super-Eddington phase, before the large energy released during the accretion swaps the gas material around the MBH and hinders a subsequent Eddington limit phase \citep{Lupi2016,Regan2019}. For simplicity, and to reduce the large number of parameters we fix $\mathcal{F}_{\rm SE} \,{=}\,0.1$. The functions $B(a)$, $C(a)$ and $D(a)$ are taken from \cite{Madau2014} and they scale with the spin of the MBH ($a$) as $B(a)\,{=}\,(0.9663 \,{-}\ 0.9292a)^{-0.5639}$, $C(a)\,{=}\,(4.627 \,{-}\ 4.445a)^{-0.5524}$ and $D(a)\,{=}\,(827.3\,{-}\,718.1a)^{-0.7060}$. 
Notice that we do not make any assumption about the spin value of the MBH since it is computed self-consistently in \LGalaxies{} after any MBHB merger and gas accretion episode \citep[see][for further details]{IzquierdoVillalba2020}\footnote{Taken into account \cite{Madau2014}, the radiative efficiency of the MBH during the super-Eddington accretion is given by $\epsilon\,{=}\,\frac{\dot{m}}{16} A(a) \left(\frac{0.985}{\dot{m}+B(a)} \,{+}\,\frac{0.015}{\dot{m}+C(a)}\right)$, being $\dot{m}\,{=}\,\dot{\rm M}_{\rm Edd}/\dot{\rm M}$.}. Finally, $\dot{\rm M}$ and $\dot{\rm M}_{\rm Edd}$ are the accretion rate and Eddington accretion rate onto the MBH, respectively. To determine $\dot{\rm M}$ we extract a random number between [0-$10^5$] distributed according to $\dot{\rm M}^{-1}$. This choice is motivated by recent theoretical works that show a power-law decline in the distribution of simulated populations of accreting MBHs for values above the Eddington-limit (see e.g \citealt{Fanidakis2012}, \citealt{Griffin2018} or \citealt{Shirakata2019})\footnote{We have also checked that the results presented here do not change significantly when the accretion rate is extracted according to a distribution following $\dot{\rm M}^{-2}$.}.\\

\subsection{Setting up the super-Eddington conditions: The \SupEdd{} model}

\begin{figure}    
	\centering  
	\includegraphics[width=0.9\columnwidth]{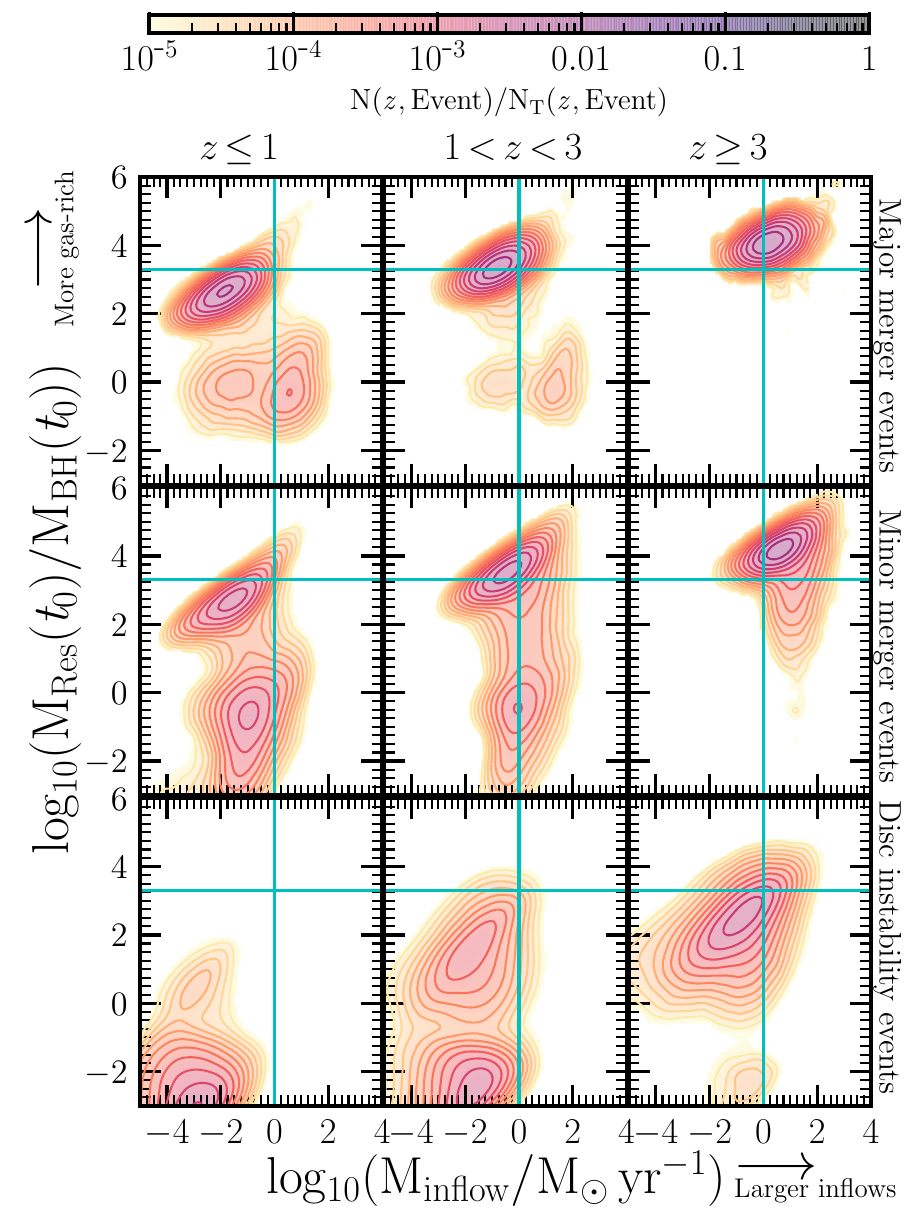}
	\caption[]{$\rm M_{infolw}$ -- $\rm M_{Res}(\mathit{t}_0)/M_{BH}(\mathit{t}_0)$ plane in three different redshifts bins according to \LGalaxies{} run on the \texttt{Millennium} merger trees. While the y-axis depicts how gar-rich is the environment around the MBH, the x-axis illustrates how powerful are the gas inflows towards the MBH. The blue horizontal and vertical lines correspond to the chosen $\mathcal{R}^{\rm th}$, and $\rm M_{inflow}^{th}$. Each row corresponds to each of the events that can trigger cold gas inflows toward the MBH: major mergers (top), minor mergers (middle) and disk instabilities (bottom). The plot has been computed by using the free parameters of the \Boosted{} model reported in Table~\ref{table:Models_Parameters}. To guide the reader, major (minor) mergers correspond to galaxy interactions which involve galaxies with baryonic masses differing by less (more) than a factor of 2.}
	\label{fig:Minflow_Distribution}
\end{figure}

The large number of possible galaxy merger histories provided by \LGalaxies{} and the \texttt{Millennium} merger trees enable us to explore which are the most common gas environments around MBHs and the typical amount of gas flowing toward the galactic center after any secular or merger process. To illustrate the interplay between these quantities, Fig.~\ref{fig:Minflow_Distribution} shows the plane $\rm M_{inflow}$ versus $\rm M_{Res}(\mathit{t}_0)/M_{BH}(\mathit{t}_0)$ for the three different events able to trigger the MBH growth: major merger, minor mergers and disk instabilities. These quantities have been computed under the assumption of a more efficient fuelling of gas onto MBHs (see the free parameters used in the third row of Table~\ref{table:Models_Parameters}). As shown, at $z\,{>}\,3$ galaxy interactions can trigger large gas inflows (${>}\,10 \msun/\rm yr$) that bury MBHs in large gas reservoirs of up to ${>}\,10^3$ times more massive than the MBH itself. 
This trend is the result of the fact that interacting high-$z$ galaxies are compact, gas-rich, and host MBHs whose mass is several orders of magnitude smaller than the cold gas component. On the other hand, disk instabilities occurring at high-$z$ do not trigger the large gas inflows shown in mergers, with rates that are typically 2 orders of magnitude smaller ($\rm 0.1\, M_{\odot}/\rm yr$). This points out that in the high-$z$ universe, galactic mergers between gas-rich systems are the unique systems that can sustain significant gas inflows to trigger potential super-critical accretion events onto MBHs. 
These results align with the findings of other works such as \cite{Pezzulli2016,Pezzulli2017} and \cite{Trinca2022} who showed, by using a semi-analytical model, that small MBH seeds at high-$z$ can increase several orders of magnitude their masses trough super-critical accretion after gas-rich galactic mergers. Furthermore, similar conclusions have been drawn from the hydrodynamical simulations of \cite{Lupi2023}, which pointed out that gas-rich environments at high-$z$ redshift can support long-lasting (tens of Myrs) super-Eddington accretion phases, speeding up the growth of intermediate-mass MBHs. Fig~\ref{fig:Minflow_Distribution} also shows that at intermediate redshifts, $1\,{<}\,z\,{<}\,3$, the picture is similar to the one seen at higher-$z$. However, there is an important decrease in events with large inflows and a rise of cases where MBHs are surrounded by small gas reservoirs. This is the consequence of the fact that a large fraction of the galaxy population already transformed its gas component into stars; as a consequence, mergers between gas-rich systems are less common than at higher redshifts. Specifically, in the case of major mergers, the peak of the $\rm M_{inflow}$ distribution is displaced down to ${\sim}\,0.1 \rm \msun/yr$, about $1.5$ orders of magnitude smaller than the higher-$z$ case. Finally, mergers and disk instabilities occurring at $z\,{<}\,1$ are very inefficient in sustaining large inflows towards the galactic nucleus, with rates that rarely exceed $0.01 \rm \msun/yr$. Such low inflows imply that MBHs are systematically embedded in small gas reservoirs whose masses can be two orders of magnitude smaller than the central MBH.\\

\begin{figure}    
\centering  
\includegraphics[width=1.\columnwidth]{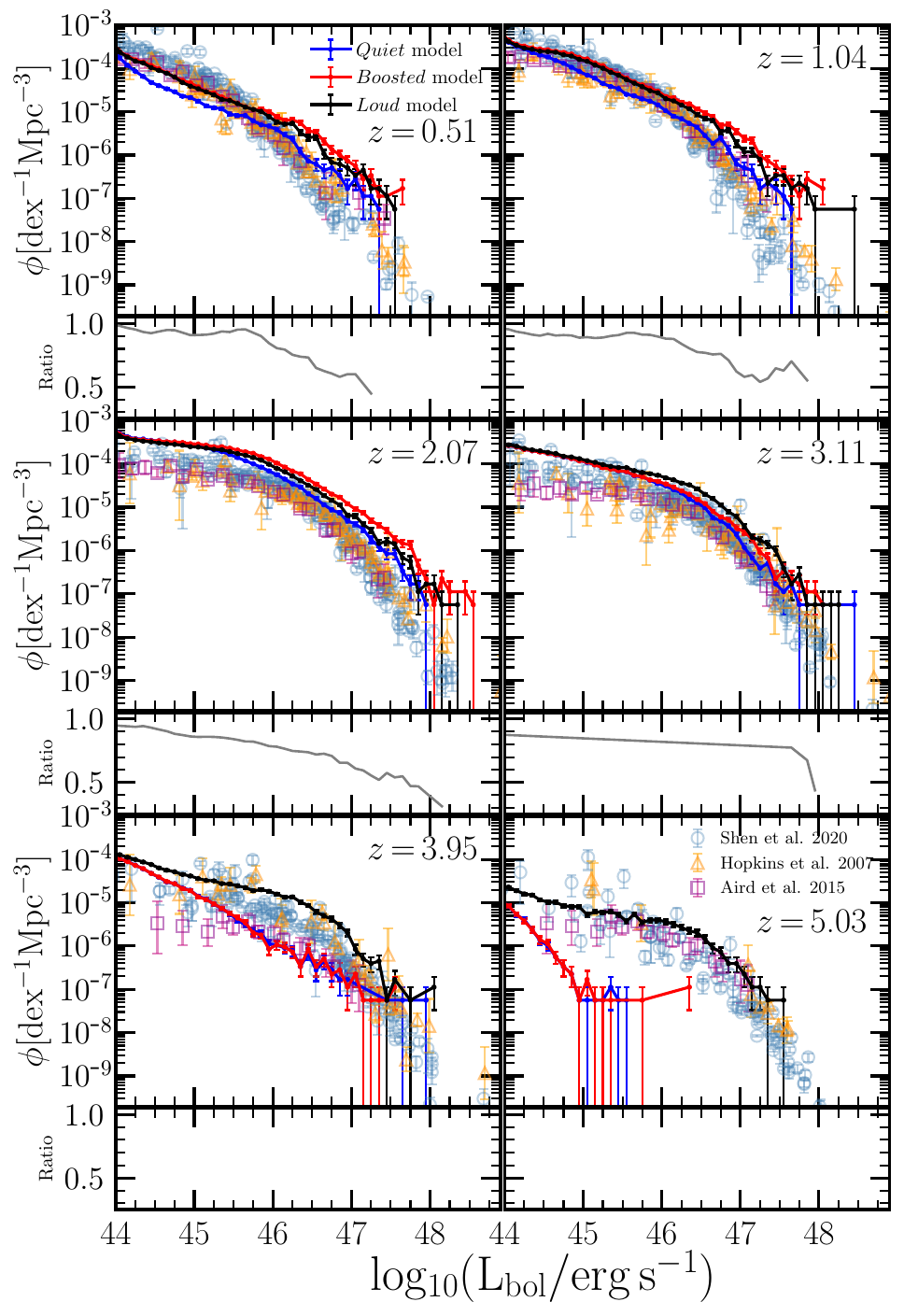}
\caption[]{Evolution of the quasar luminosity function produced by the \SupEdd{} model (black line), \Boosted{} model (red line) and \Fiducial{} model (blue line) when run on the \texttt{Millennium} merger trees. The error bars correspond to the Poissionian error. The results are compared with the observations of {\protect \cite{Hopkins2007} (orange triangles) \cite{Aird2015} (purple squares) and \cite{Shen2020} (blue circles)}. The panels below each luminosity function represent the ratio of the luminosity functions predicted by \SupEdd{} and \Boosted{} model. Notice that the two lower ones do not display any line because their ratio goes beyond the limit $1$.} 
\label{fig:SuperEddington_Luminosity_Functions}
\end{figure}

\begin{figure}
\centering  
\includegraphics[width=0.8\columnwidth]{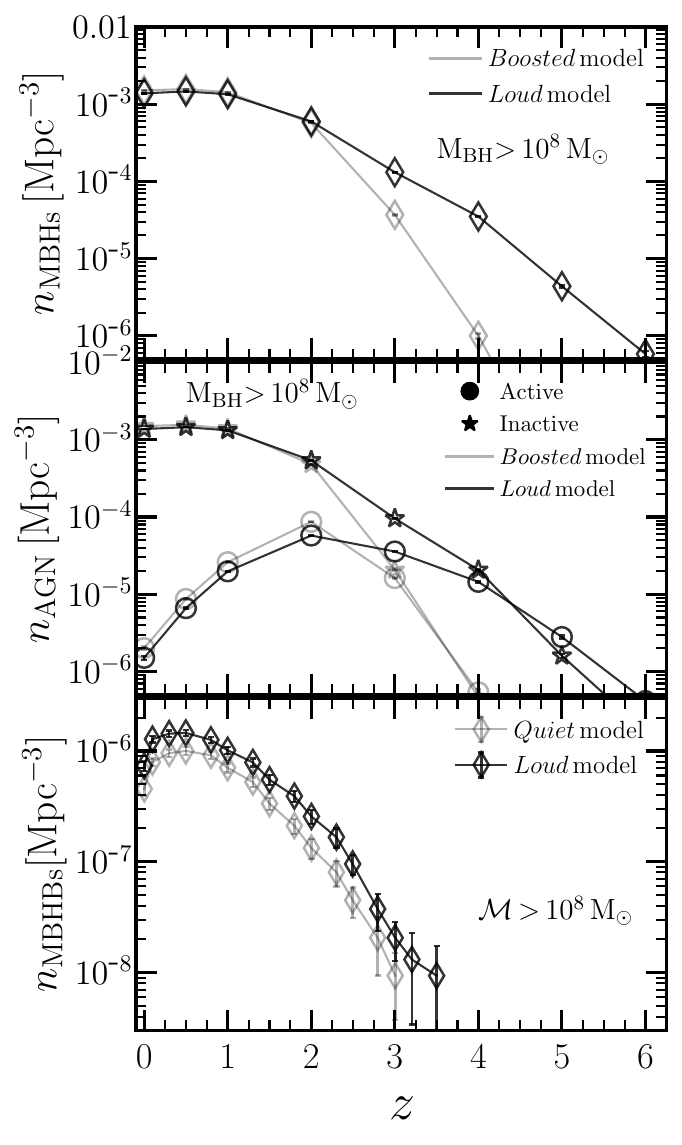}
\caption[]{\textbf{Upper panel}: Number density of MBHs with mass ${>}10^8\,\msun$ as a function of redshift, $z$, for \SupEdd{} (black) and \Boosted{} model (grey). \textbf{Middle panel}: Number density of active ($f_{\rm Edd}\,{>}\,0.01$) and inactive ($f_{\rm Edd}\,{<}\,0.01$) MBHs with mass ${>}10^8\,\msun$ as a function of redshift, $z$, for \SupEdd{} (black) and \Boosted{} model (grey). \textbf{Lower panel}: Number density of MBHBs with chirp mass $\mathcal{M}\,{>}10^8\,\msun$ as a function of redshift, $z$, for \SupEdd{} (black) model. For reference, it has been shown the results for the \Fiducial{} model (grey). In all panels, the error bars correspond to the Poisson error. We remind the reader that the comoving volume provided by the \texttt{Millennium} simulation corresponds to $\rm {\sim}\, 3.6\,{\times}\,10^8\,Mpc^3$}
\label{fig:Inactive_Active_Density_Massive_Black_Holes_and_Binaries}
\end{figure}

Taking into account the trends presented above and how the LFs evolve with different thresholds in $\rm M_{inflow}^{\rm th}$ and $\mathcal{R}^{\rm th}$ (presented in Appendix~\ref{appendix:ThresholdEffect} for the sake of brevity), we have chosen $\rm M_{inflow}\,{=}\,10\msun/yr$ and $\rm \mathcal{R}^{th}\,{=}\,2\,{\times}\,10^{3}$ as the best set of parameters to reproduce the evolution of the quasar population. The resulting model is tagged as \SupEdd{} (see Table~\ref{table:Models_Parameters}) 
and its LFs are presented in Fig.~\ref{fig:SuperEddington_Luminosity_Functions}. As shown, the match between predictions and observations improves with the new model. Specifically, the \SupEdd{} model matches the $z\,{>}\,4$ LFs and predicts at $z\,{\leq}\,2$ up to a factor of 2 less objects for any bin of luminosity larger than $\rm 10^{46} erg/s$, reducing the tension seen in the \Boosted{} case. These improvements are the result of the faster MBH population assembly which consumes most of its gas at high-$z$ and evolves quiescently in the low-$z$ Universe (see Appendix~\ref{appendix:ThresholdEffect}). This effect can be seen in the upper and middle panel of Fig~\ref{fig:Inactive_Active_Density_Massive_Black_Holes_and_Binaries} which presents the number density of MBHs with mass ${>}\,10^8\, \msun$ in the \SupEdd{} and \Boosted{} model (see the global assembly of the MBH population of the \SupEdd{} model in Appendix~\ref{appendix:Population_of_MBHs_SuperEddington}). As shown, the population of ${>}\,10^8\, \msun$ is in place much earlier in the former model than in the latter, with number densities that can be up to ${\sim}\,1\, \rm dex$ larger. Regarding the AGN activity of such population, we can see that they are mainly active ($f_{\rm Edd}\,{>}\,0.01$) at $z\,{>}\,3$ but they become rapidly inactive ($f_{\rm Edd}\,{<}\,0.01$) at $z\,{<}\,2$.\\

The better agreement between the \SupEdd{} model and the electromagnetic constraints \textit{does not imply} a small value of the sGWB amplitude which instead has a value  $A=1.8\,{\times}\,10^{-15}$, consistent with the 90\% credible interval reported by all the PTA collaborations. The reason why the sGWB is higher in the \SupEdd{} model with respect to the \Fiducial{} one can be seen in the lower panel of Fig~\ref{fig:Inactive_Active_Density_Massive_Black_Holes_and_Binaries}, which shows the number density of MBHBs with $\rm \mathcal{M}\,{>}\,10^8\msun$, i.e the systems which contribute the most to the nHz sGWB signal \citep[see][]{Sesana2008,Sesana2013,IzquierdoVillalba2021}. As we can see, these MBHBs are 
in place much earlier in the \SupEdd{} model than in the \Fiducial{} one with number densities up to 2 times larger at any redshift (especially at $z\,{<}\,1$). Finally, the implications that our new modelling has about the population of super-Eddington sources can be found in Appendix~\ref{appendix:Population_of_SuperEddington}.

\section{Constraining low-redshift LISA MBHB mergers from the nHz sGWB}
\label{sec:LISA_effects}

In this section, we study the implications of the \SupEdd{} model, featuring a sGWB of amplitude $1.8\,{\times}\,10^{-15}$ at $f\,{=}\,1 \, \rm yr^{-1}$, on the expected merger rate and electromagnetic counterpart detection of LISA MBHBs. Since we are interested in a population of MBHs up to 4 orders of magnitude smaller than the PTA one, instead of using the \texttt{Millennium} merger trees we will make use of the \texttt{Millennium-II} ones, which offer the possibility of resolving halos down to $\rm {\sim}\, 10^8\, \msun$ (see Seciton~\ref{sec:NBodySimulations}). On top of this, the high-resolution offered by \texttt{Millennium-II} enables us to seed MBHs in newly formed galaxies according to the multi-flavour seeding model of \cite{Spinoso2022}.\\

\subsection{The detection rate of LISA MBHBs}

\begin{figure}
\centering  
\includegraphics[width=1.0\columnwidth]{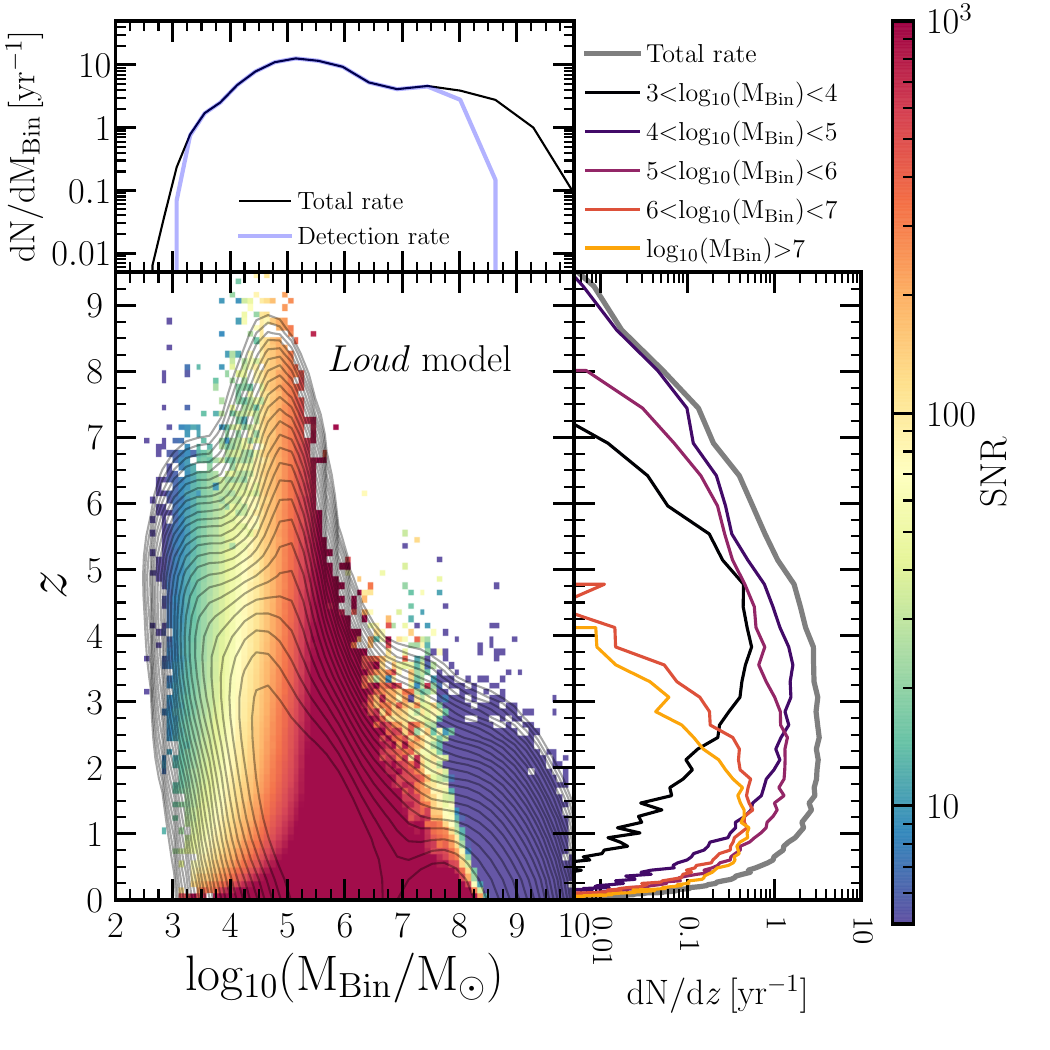}
\caption[]{Distribution of MBHB mergers predicted by the model \SupEdd{} in the plane redshift ($z$) versus total rest-frame MBHB mass ($\rm M_{Bin}$). Each pixel is encoded by the median signal-to-noise ratio (SNR). The contours represent the number of objects in the $z\,{-}\,\rm M_{Bin}$ plane. The small plots around the $z\,{-}\, \rm M_{Bin}$ plane correspond to the redshift (right panel) and source-frame mass (top panel) distribution of the differential number of MBHBs. In the right panel, the total distribution has been divided into 5 different mass bins.}
\label{fig:SNR_SE}
\end{figure}

\begin{figure}
\centering  
\includegraphics[width=1.0\columnwidth]{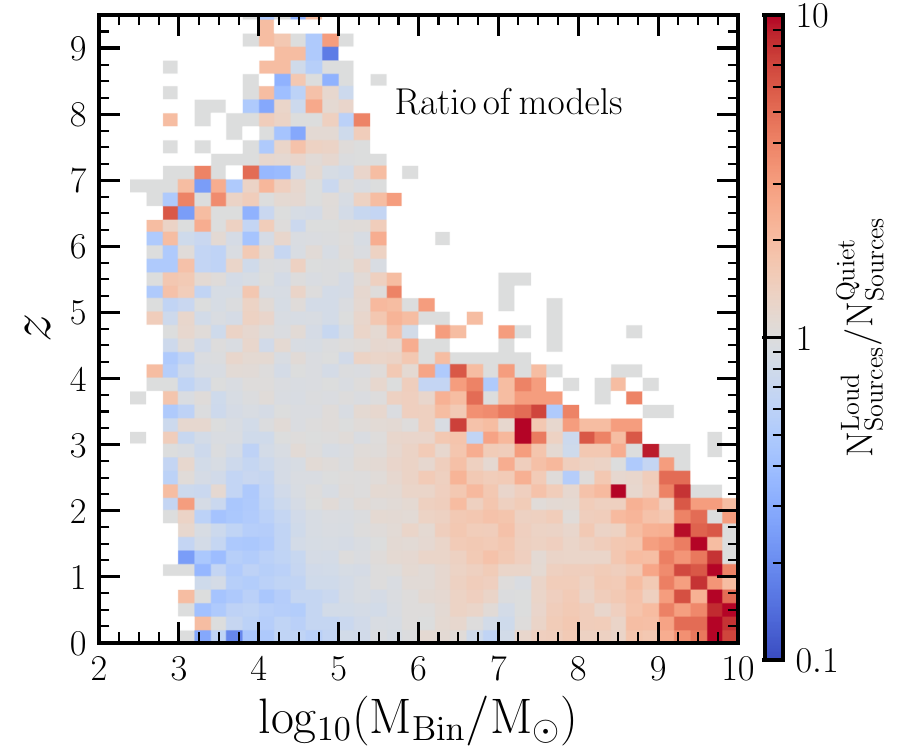}
\caption[]{Distribution of the $z\,{-}\, \rm M_{Bin}$ when each pixel is encoded by the ratio of the number of MBHB mergers in the \SupEdd{} and \Fiducial{} model.}
\label{fig:RatioModels}
\end{figure}

The redshift versus total rest-frame MBHB mass plane of merging binaries in the \SupEdd{} model is presented in Fig.~\ref{fig:SNR_SE}. 
As shown, at $\rm M_{Bin}\,{<}\,10^6\,\msun$ the \SupEdd{} model displays many mergers occurring at $z\,{>}\,3$ but the large majority of the events lie at $1\,{<}\,z\,{<}\,3$ (see similar results from hydrodynamical simulations in \citealt{Salcido2016}). For more massive systems, most mergers occur at $z\,{<}\,1$. To explore how the PTA signal affects the LISA predictions, in  Fig.~\ref{fig:RatioModels} we present the $\mathrm{M_{Bin}}\,{-}\,z$ plane where each pixel represents the ratio between the number of MBH mergers predicted by the \SupEdd{} and \Fiducial{} model. While the former displays an sGWB of $A\,{=}\,\rm 1.8{\times}\,10^{-15}$, compatible with PTA results, the latter produces a signal with $A\,{=}\,\rm 1.2{\times}\,10^{-15}$ and is in line with a large number of past theoretical works \citep[see e.g.][]{Jaffe2003,Sesana2008,Sesana2013,Roebber2016,Kelley2016,Barausse2020}. As shown, the \SupEdd{} model produces $2\,{-}\,5$ times more mergers of MBHBs with total masses ${>}\,10^6\, \msun$ than the \Fiducial{} one. These differences decrease for $\rm 10^5 \,{<}\, M_{Bin}\, {<}\,10^6\, \msun$ where the ratio varies between $0.9\,{-}\,1.5$. For the lighter systems (${<}\,10^4\,\msun$) the \SupEdd{} model displays a deficit with respect to the \Fiducial{} one. Specifically, the number of MBHB mergers at these masses can be down to 0.5 less frequent at $z\,{<}\,4$. The decrease of light MBHB mergers in favour of more massive ones is the result of an earlier and faster assembly of the MBH population in the \SupEdd{} model than in the \Fiducial{} one (see Fig.~\ref{fig:Inactive_Active_Density_Massive_Black_Holes_and_Binaries}). The different MBHB populations generated by the two models point towards different merger rates. These are summarized in Table~\ref{table:Rate_of_mergeres}. As shown, the \SupEdd{} and \Fiducial{} model displays a similar global rate of $12.7\, \rm yr^{-1}$. However, as expected from Fig.~\ref{fig:RatioModels} the differences are more evident when dividing the population by masses. Specifically, the \SupEdd{} model predicts merger rates than are ${\sim}\,1.2$ smaller than the \Fiducial{} case when MBHBs of total mass $10^{3\,{-}\,5}\,\msun$ are considered. However, it boosts by a factor of $1.5$ the coalescences involving systems with total mass ${>}\,10^5\, \msun$ ($\rm 0.8\,{-}\,4.3\, yr^{-1}$, depending on the mass).\\

Having determined the MBHB merger rate, it is important to calculate how many events are detectable by LISA. To this end, we compute the signal-to-noise ratio (SNR) of all the simulated binaries, defined as:
\begin{equation} \label{eq:SNR_equation}
    \rm SNR\,{=}\, \left(\int_{\mathit{f}_0}^{\mathit{f_f}} \left[ \frac{\mathit{h}_c(\mathit{f}'_{obs})}{\mathit{h}_n(\mathit{f}'_{obs})}\right]^2 \frac{\mathit{df}'_{obs}}{\mathit{f}'_{obs}} \right)^{1/2},
\end{equation}
where $h_{\rm n}$ corresponds to the characteristic strain noise parameterized as in \cite{Babak2021}, while $h_{\rm c}$ represents the characteristic strain amplitude of the source defined as $h_{\rm c}(f)\,{=}\,4f^2|\widetilde{h}(f)|^2$ being $\widetilde{h}(f)$ the Fourier transform of the strain signal, here computed according to the phenomenological frequency-domain gravitational waveform model \textit{PhenomC} \citep{Santamaria2010}. 
$f_0$ represents the starting frequency of inspiralling binaries, set for simplicity to the low-frequency cut-off limit of the LISA sensitivity curve ($10^{-4} \, \rm Hz$). Instead, $f_f$ is the maximum frequency of the signal set to $0.15 c^3/G (1+\mathit{z}) \,\rm {M_{Bin}}$\footnote{This specific choice about the initial and final frequencies is made since all events that are going to merge in the LISA band will generally span the whole frequency range from the cut-off up to the merger frequency.}. The resolution frequency bin, $df$, used to integrate Eq.~\ref{eq:SNR_equation} is set to $df\,{=}\, 1/T_{\rm obs}$ where $T_{\rm obs}\,{=}\,4\,\rm yr$ corresponds to the length of LISA observations. 
Finally, the effective spin of the MBHBs is taken from the predictions of \LGalaxies{} and the eccentricity is set to $0$, for simplicity.\\ 

The SNRs of the MBHBs generated with the \SupEdd{} model are depicted in Fig.~\ref{fig:SNR_SE}. Each bin of the $\rm 
M_{Bin}\,{-}\,\mathit{z}$ plane encodes the median SNR value of the MBHBs falling within it. As shown in Fig.~\ref{fig:SNR_SE}, MBHBs of ${>}\, 10^8 \, \msun$ have SNRs that rarely surpass 
a value of $10$. Conversely, MBHB mergers with masses $10^5\,{<}\,\rm 
M_{Bin}\,{<}\,10^7 \, \msun$  display large SNRs, with values that can span between $\rm 100\,{<}\,SNR\,{<}\,10^4$. For lighter systems (${<}10^4\, \msun$) the SNR decreases and the values are systematically ${<}\,100$, being the smallest for mergers of MBHBs with total mass $\rm {<}\,5\,{\times}\,10^3 \, \msun$ at $z\,{>}\,2$. Taking into account the SNR distribution of the MBHBs, Table~\ref{table:Rate_of_mergeres} presents the LISA detection rate assuming a minimum signal-to-noise ratio of $\rm 10$. As we can see, the \SupEdd{} model predicts a detection rate of $12.24$ MBHBs per year, being the systems with $\rm 10^4\,{<}\,M_{BHBs}\,{<}\,10^5\, \msun$ and $\rm 10^5\,{<}\,\rm M_{Bin}\,{<}\,10^6\, \msun$ the ones with largest rates ($\rm 5.10 \, yr^{-1}$ and $\rm 4.37 \, yr^{-1}$, respectively). 
As expected, MBHBs with masses ${>}\,10^7\, \msun$ are the most affected by the LISA sensitivity curve. In this mass range, LIA can detect only half of the total events ($\rm 0.4 \, yr^{-1}$). As discussed in Fig.~\ref{fig:RatioModels}, the main difference between the \Fiducial{} and \SupEdd{} models resides in the fact that the latter predicts larger merger events only for MBHBs of ${>}\, 10^5 \, \msun$. Since LISA is not very sensible to MBHBs of mass ${>}\,10^7\,\msun$ it implies that the overall detected rate predicted by \Fiducial{} and \SupEdd{} model does not differ much: $\rm 12.25 \, yr^{-1}$ versus $\rm 12.24 \, yr^{-1}$.\\

The results presented in this section highlight how the recent PTA signal is not significantly informative about the expected global merger rate which will be inferred from LISA observations \citep[see opposite conclusions in][]{Steinle2023}. However, under the assumption that an MBHB population is entirely responsible for the signal observed by PTAs, our detailed galaxy formation model suggests that the loud PTA signal would favour a more numerous population of MBHs and MBHBs of ${>}\,10^8\, \rm \msun$ at low-$z$ that usually expected (see Fig.~\ref{fig:Inactive_Active_Density_Massive_Black_Holes_and_Binaries} and Fig.~\ref{fig:BHMassFunction_SuperEddington}) which increases the possibilities (1.5 times larger than expected) of LISA to detect such kind of events during its lifetime. However, the physical processes involved in the efficient growth of large MBHs do not act in the same way for the smallest MBHs (${<}\,10^5\, \msun$) which in turn are the preferred targets of LISA. These objects are not able to increase their mass as fast as the most massive MBHs because the small galaxies where they are hosted are not capable of sustaining large and continuous gas inflows towards their centres after galaxy interactions. As a result, the population of low-mass MBHs (and consequently low-mass MBHBs, ${<}\,10^5\, \msun$) is only mildly affected by the modifications of MBH growth presented in this work. Thus, small changes are seen in the global population of LISA MBHB mergers (dominated by low-mass MBHs) with respect to the \Fiducial{} model.\\

\begin{table}
\begin{adjustbox}{width=\columnwidth,center}
\begin{tabular}{ccc} \hline  \hline 
\multicolumn{3}{c}{\textbf{All redshifts}} \\
                                    & Merger rate $[\rm yr^{-1}]$ &  LISA detection rate $[\rm yr^{-1}]$ \\ \hline
$\rm M_{BHBT} \, [\msun]$           & \SupEdd{} / \Fiducial{} &  \SupEdd{} / \Fiducial{} \\ \hline

No mass cut                       &      12.74 / 12.54                    &    12.28 /  12.23                                \\
$10^3\,{<}\, {\rm M_{Bin}} \, {\leq}\, 10^4 \, \msun$ &       1.40 / 1.56                    &      1.34  / 1.50                               \\
$10^4\,{<}\, {\rm M_{Bin}} \, {\leq}\, 10^5 \, \msun$ &       5.12 / 5.48                   &     5.12  / 5.48                               \\
$10^5\,{<}\, {\rm M_{Bin}} \, {\leq}\, 10^6 \, \msun$ &       4.38 / 4.17                    &    4.38  / 4.17                                \\
$10^6\,{<}\, {\rm M_{Bin}} \, {\leq}\, 10^7 \, \msun$ &       1.00 / 0.76                   &    1.00  / 0.76                                \\
${\rm M_{Bin}} \, {>}\, 10^7 \, \msun$                &       0.81 / 0.54                    &      0.44 / 0.31                                \\ \hline \hline 

\multicolumn{3}{c}{ \textbf{Low-$z$ Universe ($z\,{<}\,3$)}} \\ 

                                    & Merger rate $[\rm yr^{-1}]$ &  LISA detection rate $[\rm yr^{-1}]$ \\ \hline
$\rm M_{BHBT} \, [\msun]$           & \SupEdd{} / \Fiducial{} &  \SupEdd{} / \Fiducial{} \\ \hline

No mass cut                       &      6.55 / 6.45                    &    6.19 /  6.22                                \\
$10^3\,{<}\, {\rm M_{Bin}} \, {\leq}\, 10^4 \, \msun$ &       0.27 / 0.43                    &      0.27  / 0.43                               \\
$10^4\,{<}\, {\rm M_{Bin}} \, {\leq}\, 10^5 \, \msun$ &       1.97 / 2.32                  &     1.97  / 2.32                               \\
$10^5\,{<}\, {\rm M_{Bin}} \, {\leq}\, 10^6 \, \msun$ &       2.56 / 2.44                    &    2.56  / 2.44                               \\
$10^6\,{<}\, {\rm M_{Bin}} \, {\leq}\, 10^7 \, \msun$ &       0.93 / 0.72                    &    0.93  / 0.72                                \\
${\rm M_{Bin}} \, {>}\, 10^7 \, \msun$                &       0.77 / 0.53                    &      0.42 / 0.31                                \\ \hline \hline

\multicolumn{3}{c}{ \textbf{Low-$z$ Universe ($z\,{<}\,3$) and $\rm L_{bol}\,{>}\,10^{43}\, erg/s$}} \\ 

                                    & Merger rate $[\rm yr^{-1}]$ &  LISA detection rate $[\rm yr^{-1}]$ \\ \hline
$\rm M_{BHBT} \, [\msun]$           & \SupEdd{} / \Fiducial{} &  \SupEdd{} / \Fiducial{} \\ \hline

No mass cut                       &      2.55 / 2.19                  &    2.41 /  2.13                                \\
$10^3\,{<}\, {\rm M_{Bin}} \, {\leq}\, 10^4 \, \msun$ &       - / -                    &      - / -                               \\
$10^4\,{<}\, {\rm M_{Bin}} \, {\leq}\, 10^5 \, \msun$ &       - / -                  &     -  / -                               \\
$10^5\,{<}\, {\rm M_{Bin}} \, {\leq}\, 10^6 \, \msun$ &      1.52 / 1.48                    &    1.52  / 1.48                               \\
$10^6\,{<}\, {\rm M_{Bin}} \, {\leq}\, 10^7 \, \msun$ &       0.67 / 0.53                    &    0.67  / 0.52                                \\
${\rm M_{Bin}} \, {>}\, 10^7 \, \msun$                &       0.35 / 0.20                    &      0.22 / 0.13                                \\ \hline \hline

\end{tabular}
\end{adjustbox}
\caption{Merger rate predicted by the model (middle column) and detected by LISA ($\rm SNR\,{>}\,10$, right column) at a different total mass of the binary ($\rm M_{Bin}$). Whereas the upper part of the table depicts all the results without any cut in redshift, the middle and lower parts show the results at $z\,{<}\,3$. Furthermore, the lower part of the panel displays an extra cut in luminosity ($\rm L_{bol}\,{>}\,10^{43}\, erg/s$) referring in this way to the low-$z$ active merging MBHBs.}
\label{table:Rate_of_mergeres}
\end{table}

\subsection{Electromagnetic emission of low-$z$ LISA MBHBs}

\begin{figure}
\centering  
\includegraphics[width=1.0\columnwidth]{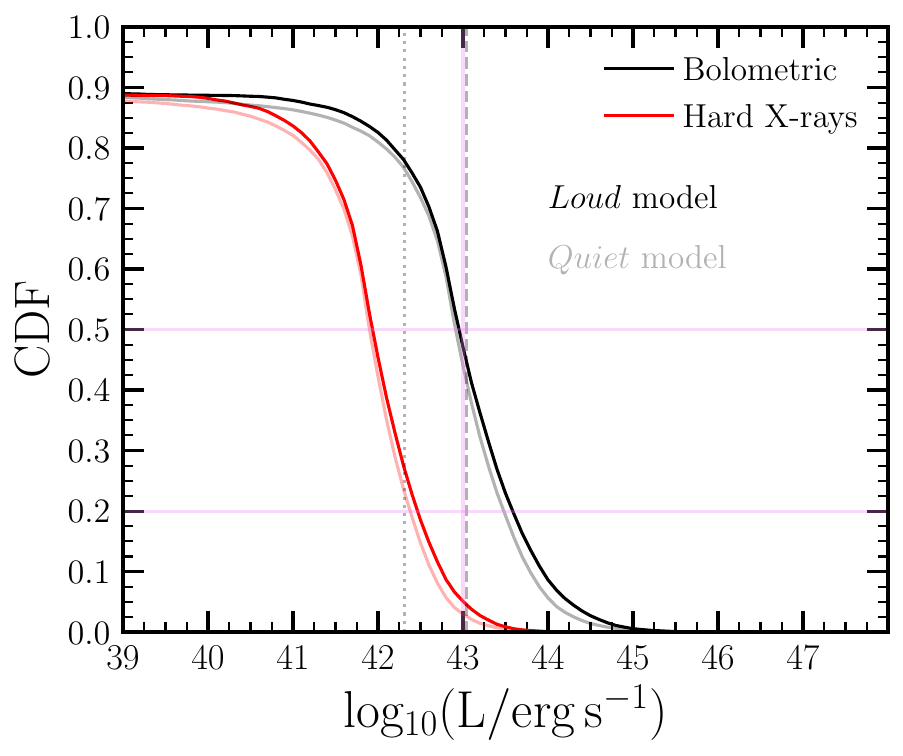}
\caption[]{Cumulative distribution function (CDF) of bolometric (black) and hard X-ray ($\rm 2\,{-}\,10\, KeV$, red) luminosity of LISA detectable ($\rm SNR\,{>}\,10$) MBHBs with masses $\rm M_{Bin}\,{>}\,10^5\,\msun$. Darker lines correspond to the \SupEdd{} model whereas the light curves depict the predictions of the \Fiducial{} one. Horizontal lines highlight the CFR values of $0.5$ and $0.2$. The vertical pink line highlights the luminosity value of $\rm 10^{43}\,erg/s$, while black dashed and dotted lines represent the minimum luminosity that a source must have respectively at $z\,{=}\,1$ and $z\,{=}\,0.5$ to be detected by Athena X-ray observatory assuming a flux limit in the hard X-ray of $2\,{\times}\,10^{-15} \, \rm erg\,s^{-1}\,cm^{-2}$ {\protect \cite[see][]{Lops2022}}.}
\label{fig:Electromagnetic_Emission}
\end{figure}

Our results suggest that the expected number of MBHB mergers detected by LISA is not significantly affected by the specific level of the nHz sGWB, as shown in Table~\ref{table:Rate_of_mergeres}. Despite that, the lower panel of Fig.~\ref{fig:RatioModels} shows that a model in agreement with current constraints on the nHz sGWB predicts that the LISA detection rate of MBHBs above $10^5\,\msun$ at $z\,{<}\,3$ could be larger compared to a model with a smaller amplitude of the sGWB. The detection and merger rates at $z\,{<}\,3$ are depicted in the middle panel of Table~\ref{table:Rate_of_mergeres}. As shown, for MBHBs of $\rm M_{Bin}\,{>}\,10^5\, \msun$ \SupEdd{} model displays up to a factor $[1.3\,{-}\,1.5]$ larger rates than the \Fiducial{} one. Since these LISA MBHBs are the ones with the brightest electromagnetic emission, this implies that during the lifetime of the LISA mission, there will be higher chances to detect the electromagnetic emission of MBHBs than previously estimated. To explore the electromagnetic detectability of LISA systems in the \SupEdd{} model, Fig.~\ref{fig:Electromagnetic_Emission} presents the cumulative distribution function of the bolometric (and hard X-rays) luminosity of detectable MBHBs with $\rm M_{\rm Bin}{>}\,10^6\,\msun$. The figure shows that these systems have a 50\% (20\%) probability of shining at $\rm L_{bol}\,{>}\,10^{43}\, erg/s$ ($\rm L_{bol}\,{>}\,10^{44}\, erg/s$). In X-rays ($2\,{-}\,10\,\rm keV$, see \citealt{Merloni2004} for the bolometric correction)\footnote{According to \cite{Merloni2004} the bolometic correction to determine the hard X-ray luminosity of an AGN is given by: $\rm log_{10}\left(L_{2\,{-}\,10\,\rm keV}/L_{bol}\right) \,{=}\, -1.69 - 0.257\mathcal{L} - 0.0078\mathcal{L}^2 + 0.0018\mathcal{L}^3$, where $\rm L_{bol}$ is the binary bolometric luminosity and $\rm \mathcal{L}\,{=}\,log_{10}(L_{bol}/L_{\sun})\,{-}\,12$.} we see similar trends, with 50\% of the MBHB
with $M_{\rm Bin}{>}\,10^5\,\msun$ shining at an X-ray luminosity ${>}\,10^{42}\, \rm erg/s$. These prospects imply that future X-ray observatories such as Athena \citep{Nandra2013} or optical surveys such as the Vera C. Rubin Observatory \citep{Ivezic2019} which feature low limiting fluxes will be able to detect the electromagnetic emission coming from several LISA MBHBs.\\

Having seen that around 50\% of the MBHBs detected by LISA will have an observable electromagnetic counterpart, it is interesting to determine the detection rate of these multi-messenger MBHBs. To this end, the lower part of Table~\ref{table:Rate_of_mergeres} summarizes for the \SupEdd{} model the detection rate of MBHBs at $z\,{<}\,3$ featuring a bolometric luminosity $\rm {>}\,10^{43}\, erg/s$. The results show that LISA will be able to detect that type of system at a rate of $\rm 2.41\,yr^{-1}$ (against the $2.13 \, \rm yr^{-1}$ displayed by the \Fiducial{} model). When the population is divided into bins of mass, we can see that MBHBs with $10^5\,{-}\,10^6\, \msun$ have a detection rate of $1.5 \, \rm yr^{-1}$ whereas MBHBs of $10^6\,{-}\,10^7\, \msun$ have a rate up a factor 2 smaller. For the case of MBHBs with $\rm M_{bin}\,{>}\,10^7\,\msun$, the \SupEdd{} model shows that the detection rate can be up to $\rm 0.22\, yr^{-1}$ (a factor 2 larger than in the \Fiducial{} model).


\section{Caveats: Multiple avenues to enlarge the nano-Hz stochastic GWB} \label{sec:Caveats}

In the recent work,  \cite{InterpretationPaperEPTA2023} explored whether state-of-the-art galaxy and MBH formation and evolution models could reproduce the nHz signal reported by the EPTA collaboration. Under the assumption that the whole signal is generated by an MBHB population, the authors showed that under standard assumptions on the MBH and MBHB evolution, SAM models generally predict a signal approximately a factor of two smaller than what detected. To overcome this limitation, it was shown that a faster dynamical evolution of MBHBs after a galaxy merger and/or a rapid and larger growth of MBHs should be required. In this work, we have explored the latter, by studying the possibility that a quick assembly of the MBH population could be caused by a larger efficiency of galactic mergers in bringing gas towards the galactic nucleus, triggering super-Eddington accretion events onto single MBHs. 
However, this approach is \textit{not the unique avenue} to reconcile a high sGWB amplitude with galaxy formation models. For instance, the dynamical models for MBHBs, generally relying on oversimplified assumptions, could be revisited. The changes in the dynamics of MBHBs would imply variations in the population of MBHBs that could enlarge the nHz sGWB without imposing any change in the population of single MBHs. Therefore, different approaches used to reach large nHz sGWB would imply different consequences for low-$z$ LISA MBHBs than those found in this work. In a future paper, we plan to revisit the dynamical and growth model of MBHBs, exploring which of the requirements involved in these processes increase the nHz signal.\\

Finally, the results presented in this work do not follow the same trends as the ones reported in \cite{Barausse2023}. By using a Press–Schechter based SAM calibrated against the PTA results, the authors showed that LISA forecasts are strongly affected by the underlying PTA signal. Specifically, \cite{Barausse2023} reported that no time delays between galactic and MBH mergers, higher accretion rates onto MBHs, and heavy MBH seeding scenarios would favour large nHz sGWBs with LISA detection rates varying between $\rm 3\,yr^{-1}$ up to $\rm 9600\,yr^{-1}$. In the case of \LGalaxies{}, a model without delays causes a decrease in the PTA signal as the result of the smaller chirp masses of MBHBs at the time of merger ($h_c \,{\propto}\,\mathcal{M}^{5/3}$, see Figure 5 of \citealt{IzquierdoVillalba2021}): he time spent by large MBHBs in the dynamical friction phase enlarges the mass of the satellite and central MBH, and the gas accretion during the hardening phase tends to make more equal mass systems, increasing in this way the chirp mass of the MBHB at merging time. The discrepancies seen between this work and \cite{Barausse2023} point out that further investigations are required to shed light on how PTA detections would impact the forecast about LISA MBHBs. Specifically, an analysis of why two SAMs that follow very accurately galaxy, MBH and MBHB evolution provide so different results would help in sharpening our knowledge about what is the main physics shaping the population of MBHBs. 


\section{Conclusions} \label{sec:Conclusions}
In this paper, we explored if the constraints on the nHz sGWB provided by the latest PTA measurements can give valuable information about the population of low-mass MBHBs (${<}\,10^7\, \msun$) that will be detectable by the LISA space-based mission. To this end, we made use of the \LGalaxies{} semi-analytical model which runs on top of the \texttt{Millennium} suite of simulations and includes detailed physical models to trace galaxy, MBH, and MBHB formation and evolution.\\ 

The starting point consisted in creating a population of MBHBs producing a nHz sGWB amplitude compatible with the latest PTA results ($A\,{=}\,1.7\,{-}\,3.2\,{\times}\,10^{-15}$). To do so, we followed \cite{InterpretationPaperEPTA2023} which pointed out that the the recent sGWB would imply a faster and larger mass growth of MBHs than usually assumed. However, \cite{IzquierdoVillalba2023} showed that raising the growth efficiency of MBHs to match a louder sGWB tends to over-predict key electromagnetic constraints such as the quasar bolometric luminosity function or the local MBH  mass function. To avoid this shortcoming and taking as reference the fiducial MBH growth model of \LGalaxies{}, we constructed a  new framework in which the increase of the galaxy merger efficiency in fuelling gas onto MBHs triggers super-Eddington accretion events.
This model, calibrated by making use of the large volume provided by the \texttt{Millennium} simulation, allowed us to create a population of MBHs which generates a sGWB amplitude of $1.8\,{\times}\,10^{-15}$ and reduces the tension with the electromagnetic constraints on the quasar luminosity function. We used this model (\SupEdd{} case) to explore the predictions for LISA MBHBs and compare them with our former fiducial model (\Fiducial{} case), which produces a smaller sGWB amplitude. To reach the range of MBHB masses targeted by LISA without important resolution limitations, we applied our models on top of the \texttt{Millennium-II} simulation (i.e the high-resolution version of \texttt{Millennium}) whose merger trees offer the possibility of tracing the cosmological assembly of galaxies and MBHs placed in halos of $[10^7\,{-}\,10^{14}]\, \msun $. The main results can be summarized as follows: 

\begin{itemize}
    \item The overall LISA detection rate of MBHBs is not significantly affected by the underlying PTA signal. The \SupEdd{} model predicts a LISA MBHB detection rate of $\rm 12.3\, \rm yr^{-1}$ whereas the \Fiducial{} model forecasts $\rm 12.2\, \rm yr^{-1}$. Therefore, under the assumption of a faster MBH assembly, our results suggest that LISA rates cannot be constrained by using the latest nHz sGWB results.\\
    
    \item The underlying PTA signal causes some differences in the mass distribution of the detected LISA MBHBs. In the \SupEdd{} model the number of $10^{3-5}\, \msun$ detectable MBHBs decreases by a factor of 1.2 with respect to what is predicted by the \Fiducial{} model. Conversely, the number of coalescence of $10^{5-7}\, \msun$ MBHBs is boosted by a factor of 1.5.\\
    
    \item The increase of detectable merging MBHBs with masses ${>}\,10^5\, \msun$ found in the \SupEdd{} model implies better prospects for multimessenger astronomy. Specifically, the model predicts that MBHBs of $10^{5-7}\, \msun$ potentially detected by LISA ($\rm SNR\,{>}\,10$) have 50\% (10\%) probability of displaying an electromagnetic emission with $\rm L_{bol}\,{>}\,10^{43} \, erg/s$ ($\rm L_{bol}\,{>}\,10^{44} \, erg/s$). Furthermore, the LISA detection rate of such type of systems at $z\,{<}\,3$ is expected to be $\rm 2.4\,yr^{-1}$.
\end{itemize}

The results listed above point out that, under the assumption of a faster MBH assembly, the PTA signal cannot constrain the expected LISA merger rate. Indeed, we have shown that a fast MBH growth can only be attained in already-massive systems, where gas is efficiently fuelled onto ${>}\,10^6\, \rm \msun$ MBHs and prompted to lead super-Eddington accretion episodes. On the contrary, MBHs with mass ${<}\,10^6\, \msun$ (i.e. the main contributors to LISA events) are not able to increase their mass as fast as the larger MBHs because the small galaxies where they reside are not capable of sustaining massive and continuous gas inflows towards their centres after galaxy interactions. As a result, the population of low-mass MBHs and MBHBs is just mildly affected by an efficient growth model with episodic super-Eddington accretion events. Consequently, the latter model only leads to small changes in the global LISA merger rate, despite producing a 1.5 louder sGWB at nHz frequencies with respect to our fiducial model. Regardless of these small differences, our results show that an efficient mass-growth model induces an increased number of merging systems with ${>}\,10^6\, \rm \msun$ that can be effectively detected by LISA. Interestingly, MBHs of these masses are also the systems which are more prone to exhibit detectable EM counterparts easily accessible with current and future astronomical facilities. Therefore, our work suggest that, if the astrophysical nature of the PTA signal is confirmed, the possibility of performing multi-messenger analysis with MBHBs could be larger than currently envisioned.\\

Finally, we stress that in this work we have included a fast assembly of the MBHs to reach the recent sGWB level reported by the PTA collaborations. Nonetheless, different dynamical models for MBHBs could result in a similar enhancement of the nHz signal without invoking any change to the whole MBH population. In an upcoming paper, we will explore this possibility by investigating the conditions leading to a rise in the nHz signal as a consequence of a modified description of the MBHB dynamics.\\ 

\section*{Acknowledgements}
We thank the B-Massive group at Milano-Bicocca University for useful discussions and comments. D.I.V. and A.S. acknowledge the financial support provided under the European Union’s H2020 ERC Consolidator Grant ``Binary Massive Black Hole Astrophysics'' (B Massive, Grant Agreement: 818691). M.C. acknowledges funding from MIUR under the grant PRIN 2017-MB8AEZ and from the INFN TEONGRAV initiative. D.S. acknowledges funding from National Key R\&D Program of China (grant no. 2018YFA0404503), the National Science Foundation of China (grant no. 12073014), the science research grants from the China Manned Space project with no. CMS-CSST2021-A05 and Tsinghua University Initiative Scientific Research Program (no. 20223080023). S.B. acknowledges support from the Spanish Ministerio de Ciencia e Innovación through project PID2021-124243NB-C21. M.B. acknowledges support provided by MUR under grant ``PNRR - Missione 4 Istruzione e Ricerca - Componente 2 Dalla Ricerca all'Impresa - Investimento 1.2 Finanziamento di progetti presentati da giovani ricercatori ID:SOE\_0163'' and by University of Milano-Bicocca under grant ``2022-NAZ-0482/B''.

%
%
\bibliographystyle{aa} 
\bibliography{references}

\appendix

\section{Effect of the thresholds} \label{appendix:ThresholdEffect}
\begin{figure*}
\centering  
\includegraphics[width=2.0\columnwidth]{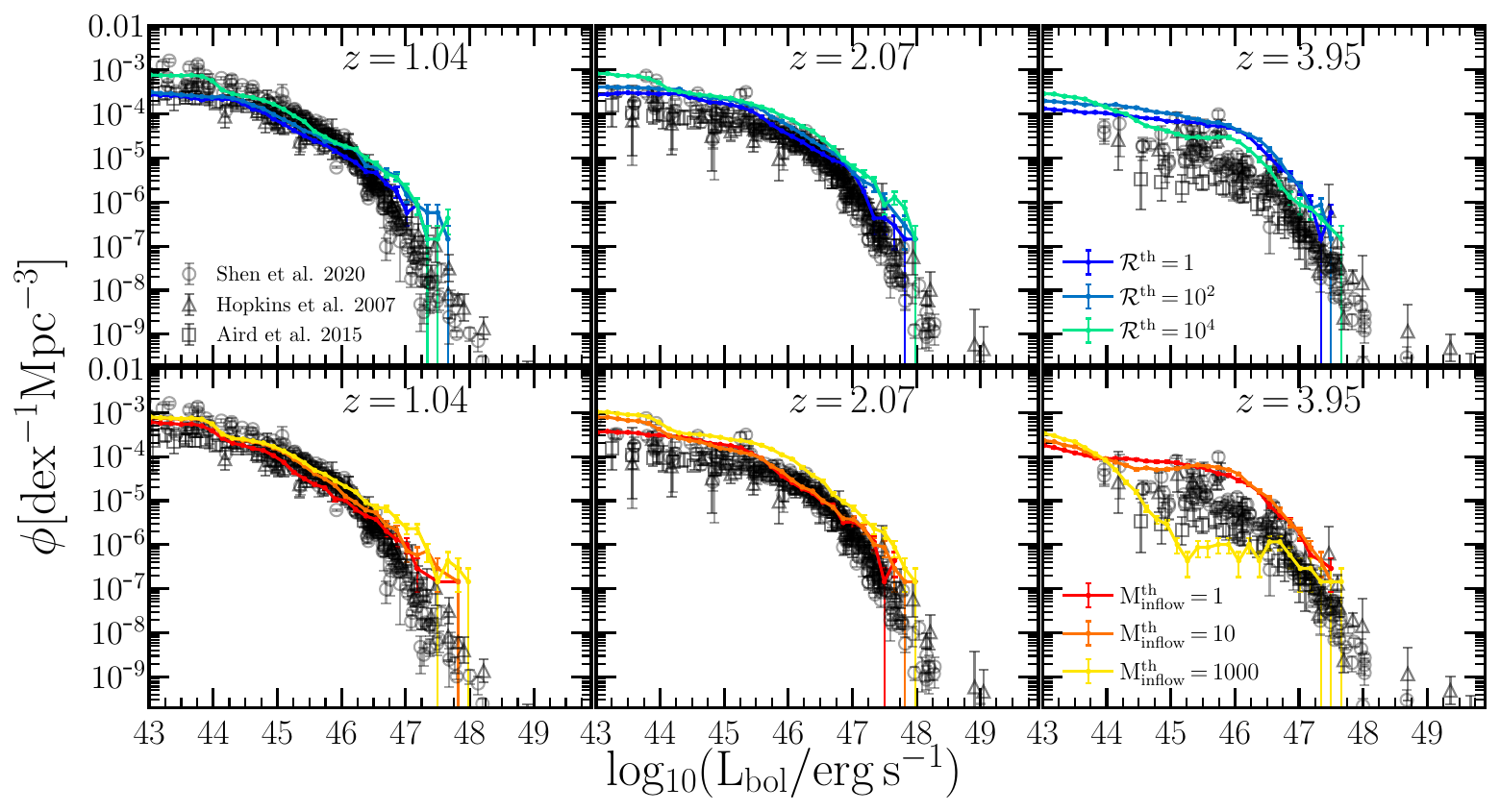}
\caption[]{AGN luminosity functions at $z\,{=}\,1.0,2.0,4.0$ for the model which includes super-Eddington events. The merger trees used correspond to the ones of \texttt{Millennium}. The error bars display the Poissionian error. The results are compared with the observations of {\protect \cite{Hopkins2007} (triangles) \cite{Aird2015} (squares) and \cite{Shen2020} (circles)}. The upper panels correspond to the predictions when varying $\rm \mathcal{R}^{th}$ and not imposing any $\rm \rm M_{inflow}^{th}$ limit. The lower ones represent the LFs of the model when changing $\rm \rm M_{inflow}^{th}$ without any $\rm \mathcal{R}^{th}$ threshold.}
\label{fig:LFs_with_SuperEddington_Effects_in_Thresholds}
\end{figure*}

In Fig.~\ref{fig:LFs_with_SuperEddington_Effects_in_Thresholds} we explore how different values of $\rm \mathcal{R}^{th}$ and $\rm M_{inflow}^{th}$ affect the number of super-critical accretion events and, thus, the evolution of the luminosity functions. Notice that when varying a threshold we do not impose any limit for the other. In this way, it is possible to marginalize the impact that each parameter. Regarding the effect of $\rm \mathcal{R}^{th}$, we can see that the smaller the value, the larger the number density of $z\,{>}\,3$ AGNs. This is because a small value of $\rm \mathcal{R}^{th}$ permits a large fraction of MBHs that are embedded in relatively gas-poor environments to trigger super-critical accretion. As a result, the population of ${>}\,10^7\, \msun$ MBHs ($\rm L_{bol}\,{>}\,10^{45}\, erg/s$) is in place earlier in the Universe, having the possibility of triggering more and brighter AGNs than in a case of a model run with large $\rm \mathcal{R}^{th}$. Interestingly, small values of $\rm \mathcal{R}^{th}$ (e.g., $1$) have an opposite effect at low-$z$, i.e. the normalization of the LFs is smaller than for the cases of large $\rm \mathcal{R}^{th}$ thresholds (e.g., $>10^2$). This smaller number density is the consequence of the faster assembly of MBHs which consumed most of their gas reservoirs at high-$z$ and thus became inactive (or quiescent) at lower redshifts. Regarding the effect of $\rm M_{inflow}^{th}$, we can see similar trends to the ones shown for $\rm \mathcal{R}^{th}$. Allowing that small inflows fuel super-critical accretion causes a faster assembly of a large fraction of the MBH population, and a rise, at $z\,{>}\,3$,  in the normalization of the LF at any bolometric luminosity. The drawback of the fast MBH assembly with small $\rm M_{inflow}^{th}$ is that at low-$z$ ($z\,{<}2$) the number of AGNs is diminished, a result of the fact that MBHs consumed more of their reservoirs at high-$z$.

\section{The population of massive black holes} \label{appendix:Population_of_MBHs_SuperEddington}

\begin{figure}
\centering  
\includegraphics[width=1.0\columnwidth]{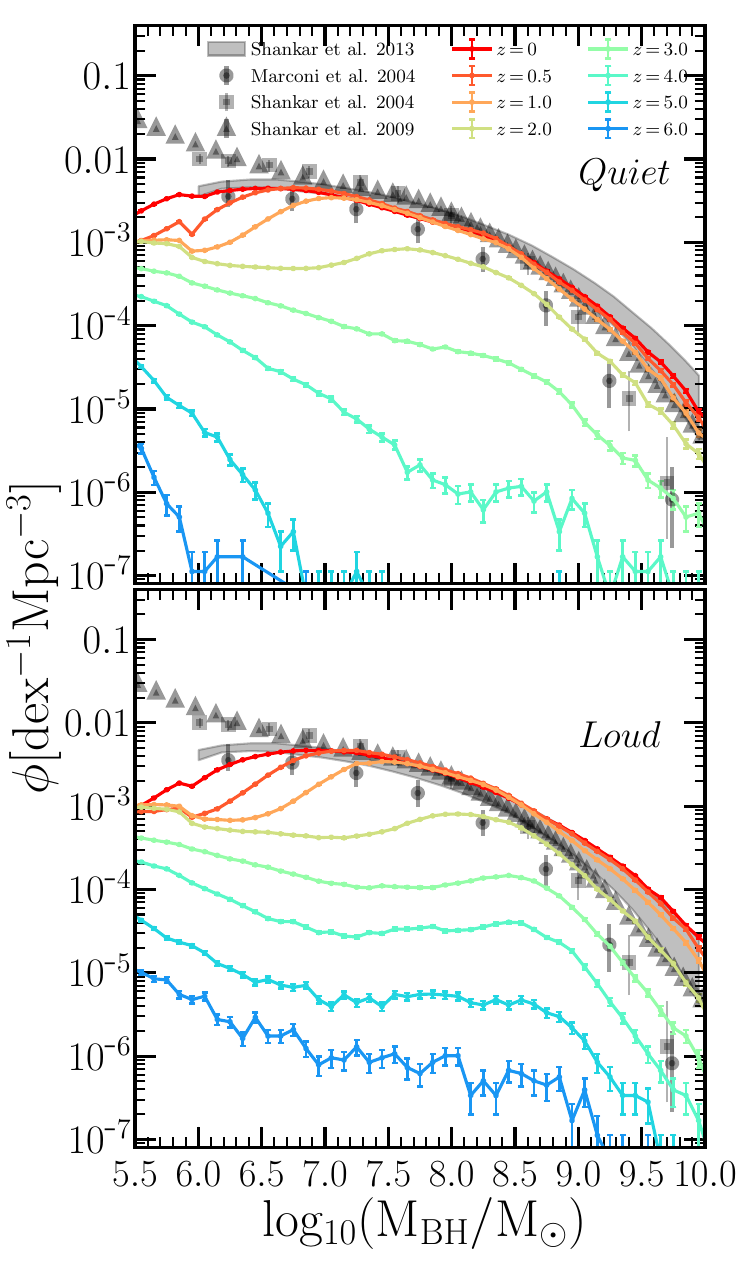}
\caption[]{Redshift evolution of the black hole mass function compared to the observational results of {\protect \cite{Marconi2004}, \cite{Shankar2004}, \cite{Shankar2009}  and \cite{Shankar2013}} for the \SupEdd{} model.}
\label{fig:BHMassFunction_SuperEddington}
\end{figure}

In this appendix, we present the population of MBHs generated by \LGalaxies{} and \texttt{Millennium} merger trees by making use of the \Fiducial{} and \SupEdd{} model. Fig.~\ref{fig:BHMassFunction_SuperEddington} depicts the evolution of the black hole mass function. As we can see, regardless of the model, the predictions at $z\,{=}\,0$ are consistent with the observational constraints provided by \cite{Marconi2004}, \cite{Shankar2004} and \cite{Shankar2013}. As described in \cite{IzquierdoVillalba2020}, the mass function displays a fast growth until $z\,{\lesssim}\,1$ moment at which it slows down and a very small evolution is seen in the massive end ($\rm M_{BH}\,{>}\,10^7\, \msun$) and the main evolutionary role is taken by the small MBH population ($\rm 10^{5.5} \,{<}\,M_{BH}\,{<}\,10^7\,\msun$). As expected, the biggest difference between  \Fiducial{} and \SupEdd{} is that the latter displays a faster assembly of the massive end of the black hole mass function. Specifically, at $z\,{\sim}\,5\,{-}\,6$ the \SupEdd{} displays a population of MBHs ($10^{7-8}\,\msun$) that is absent in the \Fiducial{} case.

\section{Implications for the MBHs: Rare events in particular hosts} \label{appendix:Population_of_SuperEddington}

\begin{figure}
\centering  
\includegraphics[width=0.8\columnwidth]{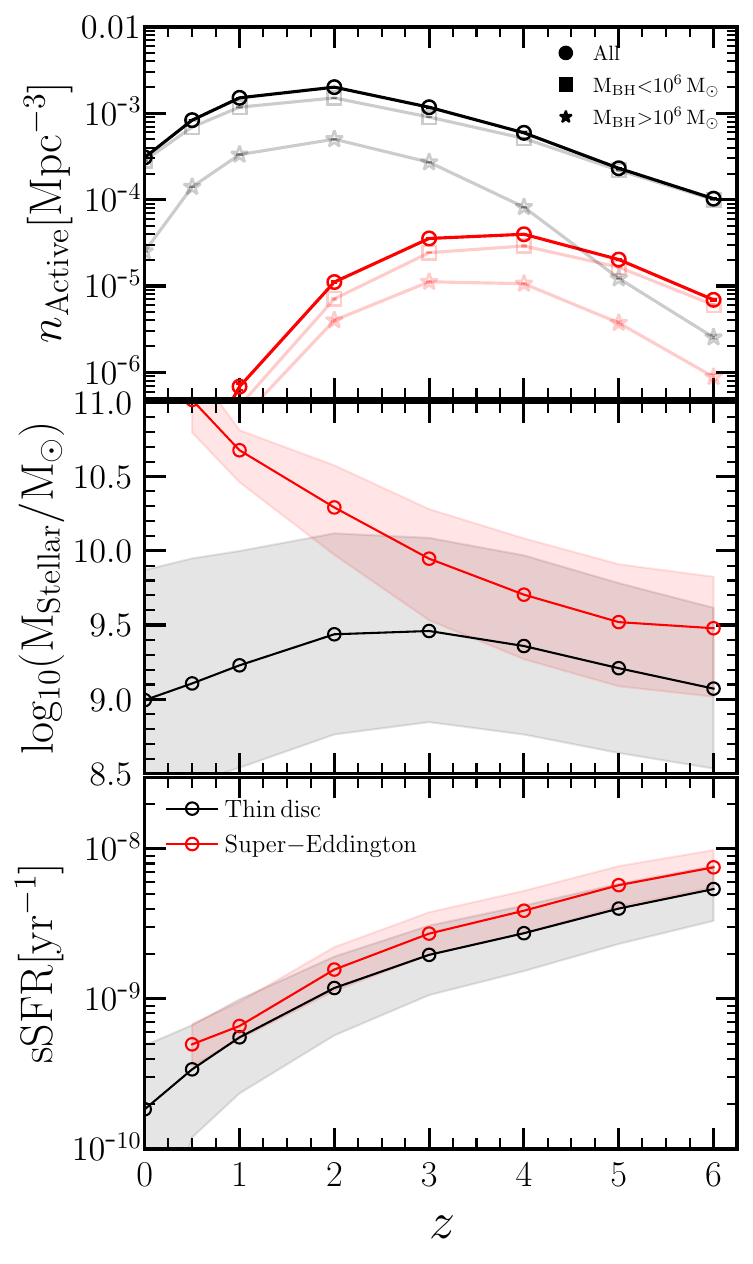}
\caption[]{Comparison between the population of MBHs accreting during the thin disk mode ($0.03\,{<}\,f_{\rm Edd}\,{\leq}\,1.0$) and super-Eddington ($f_{\rm Edd}\,{>}\,1$). The results correspond to the ones of \LGalaxies{} applied on the \texttt{Millennium} simulation. \textbf{Upper panel}: Number density of AGNs accreting at super-Eddington (red) and at thin disk (black) regimen. While circles represent the whole population of AGNs, squares, and stars correspond to AGNs triggered by MBHs with $\rm M_{BH}\,{>}\,10^6\, \msun$ and $\rm M_{BH}\,{<}\,10^6\, \msun$, respectively. \textbf{Middle and bottom panels}: Median stellar mass (specific star formation rate, sSFR) of the galaxies hosting AGNs in the super-Eddington (red) and thin disk regimen (black). The shaded area corresponds to the percentile $\rm 16^{th}\,{-}\,84^{th}$.} 
\label{fig:Number_Density_MS_sSFR_SE_Events}
\end{figure}

Since \SupEdd{} model has been calibrated using electromagnetic and GW constraints we can make predictions about the expected comoving number density of MBHs undergoing a Super-Eddington phase at different cosmological times. Fig.~\ref{fig:Number_Density_MS_sSFR_SE_Events} shows the results. As we can see, the vast majority of active MBHs are in a thin disk regime ($0.03\,{<}\,f_{\rm Edd}\,{<}\,1$), with a peak occurring at $1\,{<}\,z\,{<}\,2$ coinciding with the peak of star formation and galaxy mergers. The number density of MBHs undergoing a super-Eddington phase is up to one order of magnitude smaller and its shape displays a peak at $3\,{<}\,z\,{<}\,4$. This maximum is followed by a sharp decrease with a number densities below $\rm 10^{-5}\,Mpc^{-3}$ at $z\,{\sim}\,2$. In the upper panel of Fig.~\ref{fig:Number_Density_MS_sSFR_SE_Events} the population has been divided between massive (${>}\,10^6\,\msun$) and light (${<}\,10^6\msun$) MBHs. Interestingly, only a small fraction of light MBHs undergo a super-critical accretion. This is presumably caused by the fact that the conditions required to sustain super-Eddington episodes are not easily reachable by the small galaxies where these light MBHs reside. On the contrary, for MBHs of masses ${>}\,10^6\msun$ the requirements for super-Eddington episodes are easier to full-field provoking that the relative difference between AGNs triggered by thin disks and super-critical accretion episodes to be smaller than in the case of light MBHs.\\

Regarding the hosts of the MBHs undergoing super-Eddington accretion, Fig.~\ref{fig:Number_Density_MS_sSFR_SE_Events} shows the median stellar mass as a function of redshift. As shown, at high-$z$ ($z\,{>}\,3$) the population harboring super-Eddington MBHs does not display differences with respect to the one in a thin disk regime: the hosts display a stellar mass of $10^{8.5-9.5}\, \msun$. For lower redshifts, the galaxies where super-Eddington MBHs are placed are up to 1 dex more massive than the ones of normal AGNs ($10^9\,\msun$ versus $10^{10.5}\,\msun$). Besides stellar mass, Fig.~\ref{fig:Number_Density_MS_sSFR_SE_Events} shows the specific star formation rate (sSFR) of galaxies hosting super-Eddington and thin disk accretion. Interestingly, the former displays systematically larger values. This is the result of the fact that large galaxy inflows able to trigger super-critical accretion onto MBHs are principally related to gas-rich major mergers which, in turn, are linked with intense bursts of star formation.

\end{document}